\documentclass[a4paper,10pt,aps]{article}
\usepackage[utf8x]{inputenc}
 \usepackage[usenames,dvipsnames]{pstricks}
 \usepackage{epsfig}
 \usepackage{pst-grad} 
 \usepackage{pst-plot} 
\usepackage{amsfonts,amsthm,eucal,amssymb,amsmath}
\usepackage{graphicx}

\usepackage[normalem]{ulem}
\usepackage{color}
\definecolor{DarkBlue}{rgb}{0,0.1,0.7}

\newcommand\soutD{\bgroup\markoverwith
{\textcolor{DarkBlue}{\rule[.5ex]{2pt}{1pt}}}\ULon}
\newcommand{\Hm}[1]{\leavevmode{\marginpar{\tiny%
$\hbox to 0mm{\hspace*{-0.5mm}$\leftarrow$\hss}%
\vcenter{\vrule depth 0.1mm height 0.1mm width \the\marginparwidth}%
\hbox to
0mm{\hss$\rightarrow$\hspace*{-0.5mm}}$\\\relax\raggedright #1}}}

\title{Qualitative analysis of trapped Dirac fermions in graphene}
\author{\textsf{V\'{\i}t Jakubsk\'y, David Krej\v ci\v r\'i{}k}
 \\
{\small \textit{Nuclear Physics Institute ASCR, 
\v Re\v z, 25068, Czech Republic}}\\
{\small \textit{E-mails: jakub@ujf.cas.cz, krejcirik@ujf.cas.cz
} }
\date{\small 11 May 2014}
}

\begin{document}
\maketitle

\begin{abstract}
We  study
the confinement of Dirac fermions in graphene and in carbon nanotubes by an external magnetic field, mechanical deformations or inhomogeneities in the substrate. By applying variational principles to
the square of the Dirac operator, 
we obtain sufficient and necessary conditions for confinement of the quasi-particles. The rigorous theoretical results are illustrated on the realistic examples of the three classes of traps.    
\end{abstract}

\section{Introduction}
Collective excitations of free electrons in graphene behave like massless
Dirac fermions due to the peculiar geometry of the two-dimensional crystal, where the hexagonal lattice is assembled from two equivalent triangular sublattices \cite{Semenoff-1984-ID345}. The fact stays behind many unusual properties of graphene, e.g. the half-integer quantum Hall effect \cite{Novoselov-2005-ID389}, finite minimal conductivity \cite{Novoselov-2005-ID389}, \cite{Zhang-2005-ID435}, or visual transparency of graphene \cite{Nair-2008-ID443}. Relativistic nature of the quasi-particles makes it possible to observe phenomena native in QED in the table-top experiments. Let us mention Klein tunneling \cite{Katsnelson-2006-ID444} which is difficult to observe for elementary particles, however, it manifests in carbon nanostructures in the absence of backscattering of Dirac fermions \cite{Ando-1998-ID308}, see also \cite{Jakubsky-2011-ID20}.

Klein tunneling  challenges construction of graphene-based quantum dots and quantum wave guides as the Dirac electrons can tunnel through the electrostatic barriers. Although the quasi-particles can be confined in these systems under quite specific conditions \cite{Downing-2011-ID450}, \cite{Downing2}, \cite{Pereira-2006-ID313}, \cite{Hartmann-2010-ID45}, \cite{Hartmann-2013-ID413}, alternative ways were proposed without the use of the electrostatic field, see e.g. \cite{review} for review. In the article, we will focus on the following scenarios:

\begin{itemize}
\item
\textit{Magnetic traps:}
Inhomogeneous magnetic field can confine Dirac quasi-particles in graphene \cite{Peres-2006-ID433}, \cite{deMartino-2007-ID53}. The variety of configurations of the magnetic field leading to the confinement of Dirac fermions was considered in the literature. Let us mention e.g. the square-well barriers and point-like barriers \cite{RamezaniMasir-2008-ID440}, \cite{Dell'Anna-2009-ID12}, \cite{RamezaniMasir-2011-ID3}, Kronig-Penney-type vector potentials \cite{RamezaniMasir-2009-ID95},  or  smoothly decaying magnetic fields \cite{Roy-2012-ID415}, \cite{tarun}. Exactly solvable configurations were considered with the use of the methods of supersymmetric quantum mechanics. The standard supersymmetric techniques were used e.g. in \cite{Kuru-2009-ID442}, \cite{Milpas-2011-ID92} or \cite{Midya-2014-ID429}. Construction of solvable models and explicit formulas for their Green's functions and local densities of states were discussed  with the use of supersymmetry in \cite{Jakubsky-2012-ID85}. The wave 
guides 
created by the inhomogeneous field were considered e.g. in \cite{Myoung-2011-ID54}. 
\item
\textit{Pseudo-magnetic traps:} Influence of the mechanical deformations on the Dirac fermions can be surprisingly similar to that of the external magnetic field. They are manifested  in the form of the vector potential in the effective Dirac Hamiltonian, i.e. they give rise to a pseudo-magnetic field \cite{Kane-1997-ID103}, \cite{Suzuura-2002-ID104}, \cite{Vozmediano-2010-ID97}. Confinement of Dirac fermions caused by mechanical deformation was considered e.g. in \cite{Pereira-2009-ID199} for graphene and in \cite{Jakubsky-2012-ID85} for carbon nanotubes.
\item
\textit{Effective-mass traps:} When graphene is deposited on another crystal, the sublattice symmetry (the equivalence of the two triangular lattices in graphene) can be broken. The atoms from one sublattice are experiencing a different strength of interaction  than 
the atoms from the other sublattice. 
In the Dirac-Weyl equation, the break-down of the sublattice symmetry can be described in terms of the effective mass that can be position-dependent.  Existence of localized Dirac fermions in graphene with inhomogeneous effective mass was considered e.g. in \cite{Semenoff-2008-ID58}, \cite{mass1}, \cite{mass2}, \cite{mass3}. 

\end{itemize}

Investigation of the confinement of Dirac electrons in graphene was mostly focused on the quantitative analysis of the specific solvable configurations. In the current work, we focus our attention to the following rather general question: 
\begin{center}
\textit{Under which conditions the Dirac fermions are confined in graphene?}
\end{center} 

The article is organized as follows: In the rest of this section, we specify in detail the physical scenarios we are interested in.  In the next section, 
we find sufficient and necessary conditions 
for confinement of Dirac electrons. The main results are summarized in the form of theorem in section \ref{Theorem}. They are applied in the explicit, physically interesting, examples in section \ref{examples}. 
The last section is devoted to discussion. 

\subsection{Magnetic and pseudo-magnetic traps}
Mechanical deformation of a crystal can be described by the deformation vector $\mathbf{u(x)}$ that indicates how is the displacement of the atoms from their equilibrium positions. The effect of deformations on Dirac fermions in graphene is surprisingly similar to that of electro-magnetic field. When the crystal is subject to in-plane deformations, the stationary equation for Dirac fermions in graphene acquires the following form \cite{Suzuura-2002-ID104},  \cite{Vozmediano-2010-ID97}, \cite{Katsnelson}, 
\begin{equation}\label{eq1}
 {\sum_{j=1}^2\left[v_F\,\sigma_j\left(-i\hbar\partial_{x_j}+eA_j^{\rm mg}\right)+\gamma_0\sigma_jA^{\rm d}_j\right]\tilde{\Psi} + m\sigma_3\tilde{\Psi}=\epsilon\tilde{\Psi},}
\end{equation}
where $e$ is the elementary charge, $\sigma_a$ are the Pauli matrices.\footnote{We use standard definition of the Pauli matrices, $\sigma_1=\left(\begin{array}{cc}0&1\\1&0\end{array}\right)$, $\sigma_2=\left(\begin{array}{cc}0&-i\\i&0\end{array}\right)$, $\sigma_3=\left(\begin{array}{cc}1&0\\0&-1\end{array}\right)$. }  The mass $m$ is zero for suspended graphene but can acquire nonzero values when the graphene sheet is deposited on a substrate. The Fermi velocity  $v_F=\frac{3}{2\hbar}a_{cc}\gamma_0\approx10^6m.s^{-1}$. The hopping energy $\gamma_0=2.9eV$ and $a_{cc}=0.142\times 10^{-9}m$ is the interatomic distance in the hexagonal lattice of graphene. The vector potential $\mathbf{A^{\rm mg}}=(A_1^{\rm mg},A_2^{\rm mg})$ corresponds to the external magnetic field, whereas the vector potential $\mathbf{A^{d}}=(A_1^{\rm d},A_2^{\rm d})$ is induced by the deformation of the crystal. It is defined in terms of the deformation vector as $A_1^{\rm d}=\frac{1}{2}(\partial_{x_1}u_1-\partial_{x_2}u_2)$, $A_2^{\rm d}=-\frac{1}{2}(\partial_{x_1}u_2+\partial_{x_2}u_1)$, see \cite{Suzuura-2002-ID104}.

In the current work, we will be interested in the systems that possess translational symmetry in one direction.
We will use the units where the energy is given in the multiples of $\gamma_0$ and the length is measured in the multiples of $a_{cc}$; we make the substitutions  $x_1\mapsto \alpha x$ and $x_2\mapsto \alpha y$ in (\ref{eq1}) where $\alpha=\frac{3}{2}a_{cc}$. The vector potentials corresponding to the magnetic field and to the mechanical deformation will be of the following form 
\begin{equation}\label{vp}
 \mathbf{A^{mg}(x)}=(0,A^{\rm mg}_y(x)),\quad \mathbf{A^{d}(x)}=(0,A^{\rm d}_y(x))=(0,-\partial_{x}u_y(x)).
\end{equation}
Separating the variables in the wave function $\tilde{\Psi}(\mathbf{x})=e^{ik_y y}\Psi(x)$, we get finally
\begin{equation}\label{pmse}
 {h\Psi(x)=\left(-i\partial_x\sigma_1+k_y\sigma_2+\frac{3a_{cc}e}{2\hbar}A_y^{\rm mg}(x)\sigma_2+A^{\rm d}_y(x)\sigma_2\right)\Psi(x)+M\sigma_3\Psi(x)=E\Psi(x)}
\end{equation}
where $E=\frac{\epsilon}{\gamma_0}$ and $M=\frac{m}{\gamma_0}$.

Let us notice that both $A_y^{\rm mg}(x)$ and $A_y^{\rm d}(x)$ should be changing slowly on the interatomic distance which we will suppose to be the case. Otherwise, there could appear interaction between the states from the valleys corresponding to the two inequivalent Dirac points. In that case, the $2\times 2$ Hamiltonian in (\ref{eq1}) would not be sufficient to describe the physical situation.

When we deal with a planar system, $k_y$ can acquire any real value. The vector potential $\mathbf{A^{d}}$ in (\ref{vp}) corresponds to the deformation $\mathbf{u(x)}=(0,u_y(x))$ which is induced by unidirectional shear of the crystal. The magnetic vector potential $\mathbf{A^{mg}}$ can be induced by the parallel wires along the $y$-coordinate in a fixed distance from the crystal or by ferromagnets posed in the proximity of the graphene sheet. Let us notice in this context that the inhomogeneous magnetic fields that vary on the
scale of nanometers were realized experimentally by making structured patterns of a thin ferromagnet posed on the substrate with magnetization vector perpendicular to the surface \cite{vanRoy-1993-ID2}.

When the nanotube is considered, the $y$-coordinate is compactified and the momentum $k_y$ gets quantized. There are qualitatively two possibilities \cite{Charlier-2007-ID169}, 
\begin{equation}
 k_y=\begin{cases}\frac{n}{r}&\mbox{metallic nanotube,}\\ \frac{1}{r}\left(n\pm\frac{1}{3}\right)&\mbox{semi-conducting nanotube,}\end{cases}
\end{equation}
where $r$ is the radius of the nanotube and $n$ is an integer. In the low-energy approximation, only the values of $k_y$ are relevant  where the energy is minimal, see \cite{Charlier-2007-ID169}. For the typical nanotubes with $r\approx 15$, only the values of $k_y$ with $n=0$ are usually taken into account.  In absence of any deformations or external fields, the value of $k_y$ classifies the nanotube as metallic or semi-conducting depending on the presence of the gap between positive and negative energies. Let us notice that the vector potential $\mathbf{A^{mg}}$ in (\ref{vp}) can be generated by a circular current loop, coaxial with the nanotube. The vector potential $\mathbf{A^{d}}$ in (\ref{vp}) corresponds to the radial twist of the nanotube \cite{Kane-1997-ID103}, \cite{Jakubsky-2012-ID85}, \cite{Correa-2013-ID86}.

Twisted carbon nanotubes were observed in experiments. They appear naturally in the ropes of carbon nanotubes \cite{Clauss-1998-ID436}. They were also prepared artificially. A long carbon nanotube was anchored at its extremes to the substrate and a small metallic paddle was attached on the suspended nanotube. The paddle was then tilted by an external field \cite{Meyer-2005-ID437}. There were other prepared nanostructures based on twisted single- or multi-wall carbon nanotubes, e.g. abacus-type resonators \cite{Peng-2007-ID186} or even rotors \cite{Fennimore-2003-ID90}, see also \cite{Joselevich-2006-ID184} for a brief review.

\subsection{Effective-mass trenches}
The hexagonal boron-nitride (h-BN) has a geometrical structure that is almost identical to that of graphene, with approximately $2\%$ difference \cite{Giovannetti-2007-ID6}. The two triangular sublattices of h-BN are not equivalent as one is composed of boron whereas the other one from nitride atoms. Due to the  inequivalence of the two sublattices, there opens a gap of approximately $5.3eV$ between positive and negative energies. The relatively large energy gap prevents the framework of the Dirac equation to be applicable for description of the lowest energy bands in h-BN. 

When graphene is deposited on the substrate of h-BN, there emerges an asymmetry between the two sublattices in graphene. The carbon atoms from the first sublattice can be closer to the boron atoms whereas those from the second sublattice can be closer to the nitride atoms. In case of perfect match of the lattices, the gap of the magnitude of order $53meV$ was predicted by density functional calculations \cite{Giovannetti-2007-ID6}. Due to the small difference in the shape of the elementary cells of the two crystals, there appear periodic moir\'e patterns in the heterostructure and effective mass changes periodically \cite{Yankowitz-2012-ID4}. The gap gets smaller 
than the value anticipated in \cite{Giovannetti-2007-ID6}, however, it is non-vanishing \cite{Sachs-2011-ID7}, \cite{Song-2013-ID9}.
In \cite{Hunt-2013-ID10}, the low-energy regime of the heterostructure with the moir\'e pattern was considered in the framework of the Dirac equation with the constant effective mass.

In the current work, we will consider the heterostructure with a linear defect where the distance between graphene and h-BN increases such that there is no interaction between the two crystals. The defect resembles a straight canyon or a trench of boron-nitride covered by graphene from above. The effective mass of the Dirac fermions in graphene is position dependent, $M=M(x)$. It is nonzero on both sides of the trench, however, it vanishes identically above the trench. 
We suppose that the h-BN crystal goes away from the graphene sheet slowly enough such that the interaction does not cause inter-valley scattering.

Disregarding any deformations or external fields, the stationary equation acquires the following form
\begin{equation}\label{em}
 {h_M\Phi=(-i\partial_x\sigma_1+k_y\sigma_2)\Phi+M(x)\sigma_3\Phi=E\Phi,}
\end{equation}
where the energy and length are in the same units as in (\ref{pmse}). Let us stress that when compared to (\ref{pmse}), the vector potential in (\ref{em}) is constant but the effective mass is position dependent. The equation (\ref{em}) can be brought into the form equivalent to (\ref{pmse}) by the unitary transformation ${U=e^{i\frac{\pi}{4}\sigma_1}}$,
\begin{equation}\label{effmassh}
 {h\Psi=U^{-1}h_MU\Psi=(-i\partial_x\sigma_1+k_y\sigma_3-M(x)\sigma_2)\Psi=E\Psi,\quad \Psi=U^{-1}\Phi,}
\end{equation}
In the following section, we will focus on spectral properties of $h$. The unitary equivalence guarantees that the results will be directly applicable for $h_M$ as well.

\section{Existence of bound states with discrete eigenvalues \label{sectionVariational}}
We shall analyze the stationary equation
\begin{equation}\label{stac}
h\,\Psi(x)=E\,\Psi(x),\quad \Psi=\left(^{\psi_1}_{\psi_2}\right),\quad x\in\mathbb{R},
\end{equation}
where
\begin{equation}\label{h}
{h=\left(-i\sigma_1\partial_x+W(x)\sigma_2+M\sigma_3\right)=\left(\begin{array}{cc}M&-i(\partial_x+W(x))\\ -i(\partial_x-W(x))&-M                  \end{array}
\right).}
\end{equation}
In particular, we focus on spectral properties of the Hamiltonian $h$. The function $W(x)$  and the constant $M$ are required to be real. We also suppose $W(x)$ to be a smooth function which  is asymptotically constant, 
\begin{equation}\label{asymW}
 \lim_{x\rightarrow\pm\infty} W(x)=W_{\pm},\quad \lim_{x\rightarrow\pm\infty} W'(x)=0,
\quad |W_-|\leq |W_+|.
\end{equation}
We consider the Hamiltonian on the domain of spinors whose spin-up 
and spin-down components are square integrable on the real line together with their first derivatives. The Hamiltonian is self-adjoint on this domain.

The equation (\ref{stac}) represents a system of coupled differential equations. Its solution can be obtained rather directly as the system can be transformed into two Schr\"odinger equations for the spin-up and spin-down components of the spinor  $\Psi(x)$,
\begin{equation}\label{rce}
 h^2\Psi(x)=\left(\begin{array}{cc}-\partial_x^2+W^2{+W'}+M^2&0\\0&-\partial_x^2+W^2{-W'}+M^2\end{array}\right)\Psi(x)=E^2\Psi(x).\end{equation}
This feature of $h$ is particularly important: there are powerful tools for the spectral analysis of Schr\"odinger operators  that are, however, not applicable for Dirac Hamiltonian in general. The qualitative difference stems from the fact that the spectrum of the Schr\"odinger Hamiltonian is bounded from below, in contrast to the unbounded spectrum of the Dirac operator. For instance, the ground state energy of a Schr\"odinger operator can be estimated very well  without the explicit knowledge of the wave functions with the use of variational principles. 

In this section, we will utilize (\ref{rce}) extensively. First, we will analyze the spectrum of the Schr\"odinger operator in (\ref{rce}) 
and will find sufficient conditions 
for existence of discrete energies associated with bound states. Then we will extend these results for the Dirac Hamiltonian (\ref{h}). 

\subsection{Spectrum of the associated Schr\"odinger operator}
Let us focus on  spectral properties of the Schr\"odinger operator which is on the lower-diagonal of $h^2$.
For the sake of convenience, we rewrite it as
\begin{equation}\label{H}
H=-\partial_x^2+V,\quad V=W^2{-W'}+M^2.\end{equation} 
The potential term $V$ tends asymptotically to the constant values
\begin{equation}\label{prop}
 V_{\pm}\equiv\lim_{x\rightarrow\pm\infty} V(x)=W_{\pm}^2+M^2.
\end{equation}
Due to (\ref{asymW}), we have $V_-\leq V_+$.
The Hamiltonian $H$ is defined on the functions that are square integrable together with their first and second derivatives.

The spectrum of $H$ is a subset of positive real numbers, $\sigma(H)\subseteq[0,+\infty)$, since the Hamiltonian is defined via the square (\ref{rce}) of the self-adjoint $h$. It can be divided into two disjoint sets, the discrete spectrum, 
$\sigma_{\rm{disc}}(H)$, and the essential spectrum, $\sigma_{\rm ess}(H)$,  
\begin{equation}
 \sigma(H)=\sigma_{\rm disc}(H)\cup\sigma_{\rm ess}(H).
\end{equation}
The discrete spectrum is formed by isolated eigenvalues of finite multiplicity corresponding to the energies of bound states. The essential spectrum contains the rest; continuous spectrum (which involves the scattering states in particular),
and possible embedded eigenvalues or eigenvalues of infinite multiplicity. 
One can show that the essential spectrum of $H$ extends from $V_-$ to infinity, 
\begin{equation}\label{essential}
\sigma_{\rm ess}(H)=[V_-,\infty).
\end{equation} 
We refer to Appendix~\ref{App1}
for the proof that is based on the Neumann bracketing and Weyl's criterion \cite{R-S}. 
The exact form of the discrete spectrum cannot be obtained without the explicit knowledge of the potential $V$ and without solution of the corresponding stationary equation. However, we can specify sufficient conditions for existence of discrete eigenvalues of the Hamiltonian employing the variational principle. 

The variational minimax principle \cite{R-S}
tells us that the expectation value of energy for a normalized state from the domain $D(H)$ of $H$ is equal to or is above  the ground state energy. More precisely, there holds 
\begin{equation}\label{variational}
 \inf \sigma(H)=\inf_{\psi\in D(H)}\frac{(\psi,H\psi)}{\|\psi\|^2}=\inf_{\psi\in D(q_H)}\frac{q_H(\psi)}{\|\psi\|^2}.
\end{equation}
Here, we denoted $q_H(\psi)=\|\psi'\|^2+(\psi,V\psi)$ the energy functional (quadratic form) associated with the Hamiltonian $H$. Despite the domain $D(q_H)$ of $q_H$ being larger than that of $H$ (it consists of the states that are square integrable together with their first derivative), the last equality in (\ref{variational}) holds true as the $D(H)$ is dense in $D(q_H)$. 

The discrete spectrum of $H$ is non-empty as long as 
$\inf \sigma(H)< \inf \sigma_{\rm ess}(H)=V_-$.
A quick inspection of (\ref{variational}) reveals the necessary condition for this to happen: the function $V(x)-V_-$ has to be negative for some $x$. Otherwise, we would have $\inf \sigma(H)\geq \inf_{\psi\in D(H)}((\psi,V_-\psi)/\|\psi\|^2)= V_-$. Taking into account (\ref{essential}), it would imply absence of discrete energy levels. In the same vein, one can prove that the ground state energy lies above the minimum of the potential, i.e. $\inf\sigma(H)\geq \inf_{x\in\mathbb{R}} V(x)$.

In general, we cannot calculate 
the precise value of $\inf \sigma(H)$ in (\ref{variational}). However, we can try to find a good upper bound of $\sigma(H)$ by a clever choice of a \textit{test function} $\psi\in D(q_H)$, employing the fact that 
\begin{equation}\label{upper}
\inf\sigma(H)\leq \frac{\|\psi'\|^2+(\psi,V\psi)}{\|\psi\|^2}.\end{equation} 
Then, we can write the sufficient condition for $\sigma_{\rm disc}(H)\neq \emptyset$ in the following manner: there exists $\psi\in D(q_H)$ such that
\begin{equation}\label{ieq}\|\psi'\|^2+(\psi,(V-V_-)\psi)<0.\end{equation} 
We can see  that the negative part of $(V-V_-)$ should be ``large enough''  such that $(\psi,(V-V_-)\psi)$ compensates the positive kinetic term $\|\psi'\|^2$. The most suitable choice of the test function, giving the lowest upper bound (\ref{upper}), would depend on the actual properties of the potential $V$.  
Without this explicit knowledge (apart from the known asymptotics (\ref{prop})), let us fix the test function in the following manner
\begin{equation}\label{psiN}
  \psi\equiv\psi_N(x) := 
  \left\{\begin{array}{ccl}
     0 & \mbox{if} &\quad x \in (-\infty,-2N] \,,
     \\
     \frac{x+2N}{N} & \mbox{if}& \quad x \in (-2N,-N) \,,
     \\
     1 & \mbox{if}& \quad x \in [-N,a] \,,
     \\
     \frac{b-x}{b-a} & \mbox{if}& \quad x \in (a,b) \,,
     \\
     0 & \mbox{if}& \quad x \in [b,+\infty) \,,
  \end{array}\right.
\end{equation}
where $a$, $b$ and $N$ are real numbers and $a<b$.  
For each~$N$, $\psi_N\in D(q_H)$ so that the energy functional $q_H(\psi_N)$ is well defined. We have
\begin{equation}\label{t2}
  \|\psi'_N\|^2+(\psi_N,(V-V_-)\psi_N) = \frac{1}{N} + \frac{1}{b-a} + \int_{-2N}^{-N} (V-V_-) |\psi_N|^2 + \int_{-N}^{a} (V-V_-) 
   + \int_{a}^b (V-V_-) |\psi_N|^2 
  \,.
\end{equation}
This choice is particularly well suited for the case where $V_-<V_+$. When $V_-=V_+$, a symmetric test function obtained by substitutions
$a\mapsto N$ and $b\mapsto 2N$ 
into (\ref{psiN}) might be more convenient. We will discuss briefly the symmetric case later on. 

When (\ref{t2}) is negative in the limit $N\rightarrow\infty$, then it is possible to find  the test function $\psi_N$ with sufficiently large $N$ such that (\ref{ieq}) is satisfied. 
The first term on the right-hand side of (\ref{t2}) vanishes in the limit while the last two terms are independent of $N$. To deal with the limit of the third term effectively, let us suppose that 
\begin{center}
 $V(x)-V_-$ is either integrable 
or it is non-positive for large negative values of $x$. ($\bigstar$)
\end{center}
In the first case, we can exchange the limit with the integration\footnote{$(V-V_-) |\psi_N|^2 $ is point-wise convergent to $V-V_-$ and $|(V-V_-)\psi_N^2|\leq |V-V_-|$, which is integrable by hypothesis. Hence, we can exchange the integration and the limit due to the dominated convergence theorem.} and obtain $\lim_{N\rightarrow \infty}\int_{-2N}^{a} (V-V_-) |\psi_N|^2=\int_{-\infty}^a(V-V_-)$ where the integral is finite. In the second case (in which the latter integral is allowed to be infinite), there exists $x_0$ such that $V(x)-V_-\leq 0$ for all $x<x_0$ and we get $\int_{-2N}^{-N} (V-V_-) |\psi_N|^2\leq 0$ for $-N<x_0$. 
In both cases, we can write 
\begin{eqnarray}\label{start}
   \lim_{N \to +\infty}\left(\|\psi'_N\|^2+(\psi_N,(V-V_-)\psi_N)\right)&\leq& \int_{-\infty}^{a} (V-V_-)  
  + \frac{1}{b-a} + \int_{a}^b (V(x)-V_-) \left(\frac{b-x}{b-a}\right)^2dx \label{start1}\\
  &\leq& \int_{-\infty}^{a} (V-V_-)  
  + \frac{1}{b-a} + \frac{b-a}{3}\,\sup_{(a,b)} (V-V_-) \\
  &\leq& \int_{-\infty}^{a} (V-V_-)  
  + \frac{1}{b-a} + \frac{b-a}{3}\,\sup_{(a,\infty)} (V-V_-)
  \,\label{start3}.
\end{eqnarray} 
Hence, the condition (\ref{ieq}) is fulfilled as long as
\begin{equation}\label{tt1}
 \int_{-\infty}^{a} (V-V_-)  
  < - \frac{1}{b-a} - \frac{b-a}{3}\,\sup_{(a,\infty)} (V-V_-).
\end{equation}
Finally, taking the right-hand side of (\ref{tt1}) as a function of $b$, we can find that it acquires its minimum for $b=\frac{\sqrt{3}}{\sqrt{\sup_{(a,\infty)} (V-V_-)}}+a$. Substituting this value into (\ref{tt1}), we get  
the following result:
\begin{itemize}\item\textit{Sufficient condition no.1:\\
The system described by $H=-\partial_x^2+V(x)$, where $V$ satisfies $(\bigstar)$, possesses at least one bound state with the energy $\max\{0,\inf V\}\leq E<V_-$, provided that
\begin{equation}\label{c1}
 \int_{-\infty}^{a} (V-V_-) 
  < -\frac{2}{\sqrt{3}}\sqrt{\sup_{(a,\infty)} (V-V_-)}.
\end{equation}
}\end{itemize}

It can happen that the potential $V$ acquires large values before it converges to $V_+$, however, it changes slowly so that the absolute values of its derivative are small. In this case, the condition (\ref{c1}) can be too strong to be satisfied.  
We can find another condition that would fit better the described situation. 
Let us suppose that we can fix $a$ in (\ref{psiN}) such that $V(a)=V_-$ 
(such an $a$ always exists whenever $V_-<V_+$). 
Then we can integrate the last term in (\ref{start1}) by parts, enjoying the fact that the boundary term cancels out,
\begin{eqnarray}
\lim_{N \to +\infty} (\psi_N,(H-V_-)\psi_N) 
  &=&\int_{-\infty}^{a} (V-V_-)  
  + \frac{1}{b-a} - \int_{a}^b V'(x) \frac{(b-x)^3}{3(b-a)^2} dx\\
  &\leq& \int_{-\infty}^{a} (V-V_-)  
  + \frac{1}{b-a} -\frac{1}{12}(b-a)^2  \inf_{(a,\infty)}V'(x).
\end{eqnarray}
Fixing appropriately $b=a-\frac{6^{1/3}}{\left(\inf_{(a,\infty)}V'(x)\right)^{1/3}}$ to minimize the last two terms, we get 
\begin{itemize}\item\textit{Sufficient condition no.2:\\
Let us suppose that $V$ satisfies $(\bigstar)$ and there exists $a$ such that $V(a)=V_-$. Then, if there holds
\begin{equation}\label{c2}
 \int_{-\infty}^{a} (V-V_-)  \leq  -\frac{1}{2}\left(\frac{9}{2}\right)^{\frac{1}{3}}\left(\inf_{(a,\infty)}V'\right)^{\frac{1}{3}},
\end{equation}
the Hamiltonian $H=-\partial_x^2+V(x)$ has at least one bound state with the positive energy that lies below the threshold  $V_-$ of the essential spectrum. }
\end{itemize}

Finally, when $V_-=V_+$, it can be more convenient to work with a symmetric test function which is defined by (\ref{psiN}) after the substitutions $a= N$ and $b= 2N$. Let us still suppose that $V-V_-$ is either integrable or negative for all $|x|>x_0$ for some $x_0>0$. Then we get $\lim_{N \to +\infty} (\psi_N,(H-V_-)\psi_N) 
 \leq \int_{-\infty}^{\infty} (V-V_-) $ instead of (\ref{start3}). 
In this way, the following sufficient condition for $\sigma_\mathrm{disc}(H)\neq \emptyset$ is obtained:
\begin{itemize}
\item \textit{  Sufficient condition no.3:\\
The Hamiltonian $H=-\partial_x^2+V(x)$ has non-empty discrete spectrum provided that 
\begin{equation}\label{c3}
 \int_{-\infty}^{\infty} (V-V_-) 
  < 0.
\end{equation}
}
\end{itemize}

\subsection{Discrete energy levels of $h$\label{Theorem}}
Let us suppose that we have $\psi_2$ that solves $H\psi_2=E^2\psi_2$ where $E^2\neq M^2$. Then we can construct two eigenstates of $h$
\begin{equation}\label{hPsi}
 {\Psi_{\pm } =\left(\begin{array}{c}\frac{-i(\partial_x+W(x))}{\pm |E|-M}\psi_2\\\psi_2\end{array}\right),}
\end{equation}
that solve the equation
\begin{equation}\label{hstac}
 h\Psi_{\pm}=\pm|E|\Psi_{\pm }.
\end{equation}
The latter formula implies that for any discrete non-zero energy $E^2\neq M^2$ of $H$, there are two discrete energy levels $E$ and $-E$ in the spectrum of $h$. Indeed, when $\psi_2$ is a bound state of $H$ (which is square integrable together with its first and second derivative), then the upper components of $\Psi_{\pm}$ preserve the square integrability including its first derivative. Hence, (\ref{hPsi}) represents bound states of $h$. With the use of (\ref{hPsi}), one can show that this spectral symmetry ($E\in\sigma(h)\Rightarrow -E\in\sigma(h)$) is not exclusive for the discrete spectrum but holds true for all energy levels (except of $|E|^2=M^2$), see the end of appendix A for more details. 

The formula (\ref{hPsi}) fails to provide eigenvectors of $h$ as long as $|E|=M$. In that case, we can find the following explicit solutions $\Psi_{\pm}=(\psi_1,\psi_2)^t$ of $h\Psi_{\pm}=\pm M\Psi_{\pm}$,
\begin{eqnarray}\label{Menergy1}
 \psi_2&=& \beta e^{{-\int^x_{c_1}W(t)dt}}, \quad 
 \psi_1=e^{{\int^x_{c_1}W(t)dt}}\left(2{iM}\beta \int^x_{c_2}e^{{-2\int^s_{c_3} W(t)dt}}ds+\alpha\right),\quad E=M,\\
\psi_1&=& \beta e^{{\int^x_{c_1}W(t)dt}}, \quad 
 \psi_2=e^{-\int^x_{c_1}W(t)dt}\left(-2iM\beta\int^x_{c_2}e^{2\int^s_{c_3} W(t)dt}ds+\alpha\right),\quad E=-M,\label{Menergy2} 
\end{eqnarray}
where $\alpha\in \mathbb{C}$, $\beta\in\mathbb{C}$ and $c_1$, $c_2$, $c_3$ are arbitrary real constants. Square integrability of the wave functions depends on the values of $W_-$, $W_+$ and $M$. As long as $W_-W_+> 0$, the wave functions are not square integrable as  $\psi_2$ in (\ref{Menergy1}) and $\psi_1$ in (\ref{Menergy2}) diverge. When $W_-W_+<0$, one of (31), (32) is square integrable for $\beta=0$. It follows from the asymptotic behavior of $W$ that (31) and (32) cease to be square integrable for $W_-W_+<0$ when $M\neq 0$ and $\beta\neq0$.

Before we summarize our findings, let us notice that the results of the preceding subsection are based on the spectral analysis of the the Hamiltonian (\ref{H}). However, we could begin equally well with the operator $-\partial_x^2+W^2+W'+M^2$ corresponding to the upper diagonal of (\ref{rce}). Then the  results would be slightly modified, just by changing the sign of $W'(x)$ in the definition (\ref{H}) of $V$. The eigenvectors of $h$ would be defined by ${\Psi_{\pm } =\left(\psi_1,\frac{-i(\partial_x-W(x))}{\pm |E|-M}\psi_1\right)^t}$ where $\psi_1$ would be required to be square integrable together with its first and second derivative. 

Now, we can summarize the sufficient and necessary conditions for $\sigma_{\rm disc}(h)\neq \emptyset$ in the following statement:\\
 
\noindent 
{\it \textbf{Theorem:} Let us consider the Hamiltonian
\begin{equation}\label{th}
h=-i\sigma_1\partial_x+W\sigma_2+M\sigma_3
\end{equation}
on the space of integrable function over the real line where $W$ is a smooth real function and $M$ is a real constant. We suppose that $W(x)$ is asymptotically constant,  $\lim_{x\rightarrow\pm\infty}W(x)=W_{\pm}$ and $|W_-|\leq |W_+|$, $|W_{\pm}|<\infty$, and that $W'$ vanishes at infinity.
Let $W^2-W_-^2+\epsilon W'$  be either integrable or there exists $x_0$ 
such that $W(x)^2-W_-^2\leq0$ for all $x<x_0$ and $\epsilon\in\{-1,1\}$.
Then 

\begin{enumerate}
\item There holds
\begin{equation}\label{symmetry}
 E\in\sigma(h)\Leftrightarrow -E\in\sigma(h)\quad\mbox{for all}\quad |E|\neq M.
\end{equation}
and the essential spectrum of $h$ is formed by two disjoint intervals
\begin{equation}\label{essp}
\sigma_{\rm ess}(h)=\left(-\infty,-\sqrt{W_-^2+M^2}\right]\cup \left[\sqrt{W_-^2+M^2},+\infty\right). 
\end{equation}

\item If there holds that
\begin{itemize}
\item there exists $a\in\mathbb{R}$ such that
\begin{equation}\label{cc1}
\int_{-\infty}^a\left(W^2-W_-^2\right)<-\frac{2}{\sqrt{3}}\sqrt{\sup_{(a,\infty)}\left(W^2+\epsilon W'-W_-^2\right)}-\epsilon (W(a)-W_-)    
\end{equation}
or\\
\item there exists $a\in\mathbb{R}$ such that $W(a)^2+W'(a)=0$ and
\begin{equation}\label{cc2}
 \int_{-\infty}^a\left(W^2-W_-^2\right)< -\frac{1}{2}\left(\frac{9}{2}\right)^{\frac{1}{3}}\left(\max_{(a,\infty)}(W^2+\epsilon W')'\right)^{\frac{1}{3}}- \epsilon (W(a)-W_-),    
\end{equation}
or\\
\item there holds 
\begin{equation}\label{cc3}
 \int_{-\infty}^{\infty} (W^2-W_-^2) 
  < 0,
\end{equation}
\end{itemize}
then the Hamiltonian (\ref{th}) has at least one bound state with the energy  
\begin{equation}
\label{energy}E\in\left(-\sqrt{W_-^2+M^2},\sqrt{W_-^2+M^2}\right).\end{equation}

\item There is a zero mode in the system (i.e. $E_0\equiv 0\in\sigma(h)$) if and only if $W_-W_+<0$ and $M=0$.
\item When both $W^2\pm W'-W_-$ are non-negative, then there are no discrete eigenvalues in the spectrum of $h$. 
When one of $W^2\pm W'-W_-$ is non-negative and $W_-W_+>0$, then $\sigma_{\rm disc}(h)=\emptyset$.\\

\end{enumerate}}

The relations (\ref{cc1}), (\ref{cc2}) and (\ref{cc3}) are direct consequences of (\ref{c1}), (\ref{c2}) and (\ref{c3}), respectively. Let us stress again that they represent a sufficient but not necessary condition for existence of discrete energies. The necessary condition is presented in the fourth statement. It is based on the fact that if $V-V_-$ is non-negative, then the spectrum of $H$ satisfies $\sigma(H)=[V_-,\infty)$. When $W_-W_+>0$, then the normalizable spinors have always nonzero spin-up 
and spin-down components. For $M=0$, the zero mode can have vanishing spin-up or spin-down component. See also the corresponding comments below (\ref{variational}) and below (\ref{Menergy2}).

The integral conditions (\ref{cc1})--(\ref{cc2}) imply that when $W^2-W_-^2$ converges to zero from below slowly enough, there always exists a bound state. Indeed, when $W(x)^2-W_-^2\leq -|x|^{\varepsilon}$ for $\varepsilon\in[-1,0)$ and large $|x|$, the integrals in (\ref{cc1})--(\ref{cc2}) are infinite and, hence, the corresponding inequalities are satisfied.

The conditions (\ref{cc1}), (\ref{cc2}) and (\ref{cc3}) are rather qualitative and do not provide any quantitative information about the discrete spectrum. However, we can use them indirectly to analyze the gap between positive and negative energies of $h$. Due to (\ref{symmetry}), the gap $\Delta(h)$  for $-M \notin \sigma_{\rm disc}(h)$, $M\notin \sigma_{\rm disc}(h)$ can be defined as
\begin{equation}
 \Delta(h)=2|\inf\sigma(h)|.
\end{equation}
The formula gives $\Delta(h)=0$ when zero is in the spectrum. The situation when the zero-mode is missing in the system is more interesting. 
We can use (\ref{rce}) to find the lower and upper estimate of the gap. We know that for the Schr\"odinger Hamiltonian $H$ there holds
$\inf\sigma(H)\geq\inf V$. We also know that $\sigma(H)$ contains non-negative values only, $\sigma(H)\subset [0,\infty)$. 
As $V$ can acquire negative values, we get $\inf\sigma(H)\geq \max\{0,\inf V\}$. Considering the operator $h$, it is convenient to define  
\begin{equation}\label{V0}
V_0=\min\left\{\inf_{x\in\mathbb{R}} (W(x)^2-W'(x)+M^2),\inf_{x\in\mathbb{R}} (W(x)^2+W'(x)+M^2)\right\}.\end{equation} 
Then $\sqrt{\max\{V_0,0\}}$ is less then or equal to $|\inf\sigma(h)|$.  Additionally, we can use (\ref{upper}) for the upper bound of the gap. Let us suppose that one of (\ref{c1}), (\ref{c2}) or (\ref{c3}) is satisfied. 
Then it is guaranteed that for some (large) $N$, (\ref{upper}) lies below $V_-=W_-^2+M^2$. Hence, we have the following upper and lower estimate of the spectral gap,
\begin{equation}\label{gapapprox}
 2\sqrt{\max\{V_0,0\}}\leq\Delta(h)\leq2\sqrt{\frac{\|\psi'_{N}\|^2+(\psi_{N},(W^2+\epsilon W'+M^2)\psi_{N})}{\|\psi_{N}\|^2}}.
\end{equation}
Here we have freedom to select $\epsilon\in\{-1,1\}$ to get a better upper bound. 

\section{Trapping of Dirac electrons: Examples\label{examples}}

\subsection{Radially twisted carbon nanotubes}
Let us consider a single-wall carbon nanotube which is subject to the radial twist. The carbon atoms are shifted from their equilibrium position perpendicularly to the axis of the nanotube while keeping the tubular shape of the nanotube. Identifying $x$-coordinate with the axis of the nanotube, the atoms are shifted along $y$-axis and the displacement vector is $\mathbf{u}=(0,u_y(x))$. Then the vector potential induced by the deformation acquires the form required in (\ref{vp}),  $\mathbf{A^{d}(x)}=(0,A^{\rm d}_y(x))$. Recalling (\ref{pmse}), we identify $W(x)$ in (\ref{th}) as
\begin{equation}\label{Wtube}
W(x)=A^{\rm d}_y(x)+k_y,\quad M=0 \end{equation}
We suppose that $W$ is asymptotically constant and nonzero.

\begin{figure}[ht]

\centering
\begin{tabular}{cc}
\includegraphics[scale=.5]{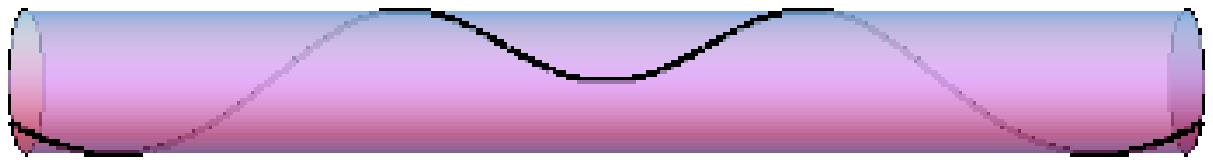}&\includegraphics[scale=.5]{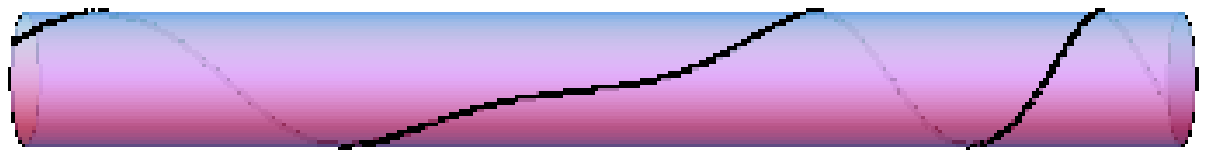}\\
\includegraphics[scale=.6]{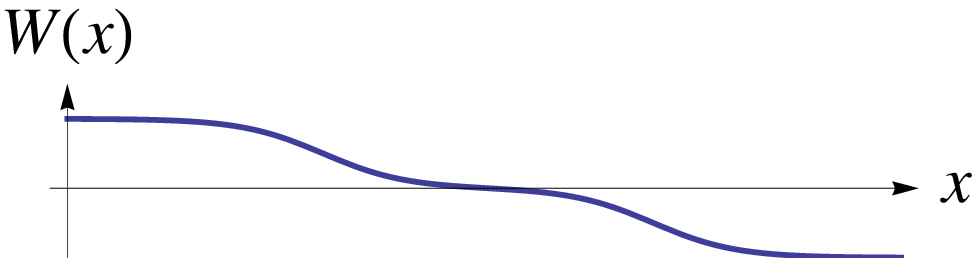}&\includegraphics[scale=.6]{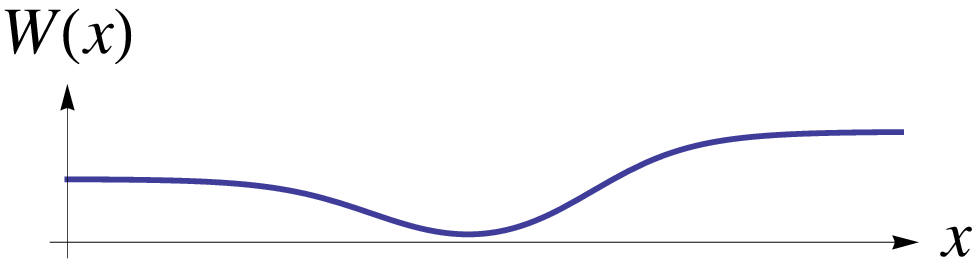}\\
\end{tabular}
\caption{Metallic carbon nanotubes ($k_y=0$) with different orientations of the twist in the ends and the corresponding vector potential $W$ below. In the nanotube without twist, the black line would be straight.\label{fig1}}
\label{tab:gt}
\end{figure}

Using the theorem presented in the preceding section, we can make interesting qualitative predictions about existence of discrete energy levels. First, when the metallic nanotube ($k_y=0$) is twisted clock-wise on one end and anti-clock-wise on the other end (see Figure~\ref{fig1} for illustration), there is a bound state with zero energy. Indeed, in this case we get $W_+W_-<0$. When the twist orientation is the same on both sides of the nanotube, the zero-mode can still exist provided that $k_y\neq 0$ compensates one of the asymptotic values of $A^{\rm d}_y$ such that we have $W_+W_-<0$ again. These are examples of physical realization of a one-dimensional domain wall \cite{Semenoff-2008-ID58}.

When $W_-W_+>0$, existence of the discrete energies is less obvious and depends on the more specific properties of the twist. First, let us consider metallic nanotubes ($k_y=0$) and suppose that $W'(x)>0$ for all $x$, i.e. the angle of the twist increases monotonically between its asymptotic values $W_-$ and $W_+$. Then the discrete energy levels are absent.  Indeed, we have $W^2-W^2_-+ W'\geq0$. 
Then the last statement of the theorem tells us that there are no discrete energy levels in the system. 

When $W(x)$ is not monotone, the finer characteristics of the twist are decisive for existence of the discrete energy levels. 
To illustrate the situation, let us consider the model where the twist gets locally damped from its asymptotic value $W_-$ and then rises to $W_+$ (see Figure \ref{fig1} (Right) for illustration),
\begin{equation}
 W(x)=\alpha_0-\alpha_1\tanh(\beta x)+\alpha_2\tanh(\beta (x-\gamma)),
\end{equation}
where $\beta$, $\gamma$ and $\alpha_j$ for $j=0,1,2$ are positive real parameters and $\alpha_0\geq\alpha_2+\alpha_1>\alpha_2\geq \alpha_1>0$. Hence, $W(x)$ is positive for all $x$.
It corresponds to the twist that is asymptotically constant,
\begin{equation}
 W_-=\alpha_0+\alpha_1-\alpha_2,\quad W_+=\alpha_0-\alpha_1+\alpha_2,\quad W_-W_+>0.
\end{equation}

It is convenient to utilize the relation (\ref{cc1}) for analysis of the discrete energies. It is possible to compute the integral 
$\int_{-\infty}^a(W^2-W_-^2)$ analytically for any $a$. However, we prefer to proceed in a simpler manner by fitting the potential $W^2$ by a square well potential $V_T$,
\begin{equation}
 V_T=\left\{\begin{array}{lll}W_-^2,& x\in(-\infty,c),&\\
		  W^2(a),& x\in[c,a],& 0\leq c<a\leq \gamma,\\
		  W_+^2,&x\in(a,\infty),&
           \end{array}
\right.
\end{equation}
where $c$ and $a$ are fixed such that $W_->W(a)\geq W(x)$ for $x\in (c,a)$. 
Then we have
\begin{equation}
 \int_{-\infty}^a(W^2-W_-^2)\leq \int_{-\infty}^a(V_T-W_-^2) =-(a-c)\left(W_-^2-W^2(a)\right)
\end{equation}
which is negative.
Considering the right-hand side of (\ref{cc1}) with $\epsilon=1$, we can write
\begin{eqnarray}
 \sup_{(a,\infty)}(W^2- W_-^2+W')
\leq W^2_{+}-W_-^2+\beta \alpha_2=4\alpha_0(\alpha_2-\alpha_1)+\beta \alpha_2,
\end{eqnarray}
where we used $W'=\beta(-\alpha_1 {\rm sech}^2\beta x+\alpha_2 {\rm sech}^2\beta(x-\gamma))\leq \beta  \alpha_2$ for all $x$.
Using the inequalities in (\ref{cc1}), the sufficient condition for existence of discrete energy levels can be written in the following form
\begin{equation}\label{estimate}
-(a-c)\left(W_-^2- W^2(a)\right)-(W_-- W(a))\leq -\frac{2}{\sqrt{3}}\sqrt{4\alpha_0(\alpha_2-\alpha_1)+\beta \alpha_2}.
\end{equation}
When the inequality is satisfied, there are discrete energies in the system. The formula specifies what the sufficient length and strength of the damping (reflected by $(a-c)$ and $W_--W(a)$, respectively) are
for confinement of the Dirac fermion. 

Instead of making the estimate of the discrete energy, let us mention that the model represented by (\ref{Wtube}) is reflectionless for a one-parameter family of the constants $\alpha_j=\alpha_j(\lambda)$, $\beta=\beta(\lambda)$ and $\gamma=\gamma(\lambda)$ where $\lambda$ is the absolute value of the discrete energy level, see \cite{Jakubsky-2012-ID85} for more details.

\subsection{Boron-nitride trenches covered by graphene}
Let us consider a composite crystal where graphene is posed on the layer of h-BN. The interaction between the crystals gives rise to a constant effective mass \cite{Hunt-2013-ID10}. We suppose that there is a linear trench in boron-nitride which causes inhomogeneity of effective mass $M(x)$ in (\ref{effmassh}). We suppose that lattices of the two crystals are perfectly matched at the edges of the trench such that each carbon atom is either above the boron or above the nitride atom. In this case, the sublattice $A$ of graphene is paired with one type of atoms (e.g.. with boron), while the sublattice $B$ is paired with the other type of atoms (with nitride).  

\begin{figure}[ht]

\centering
\begin{tabular}{cc}
\includegraphics[scale=.55]{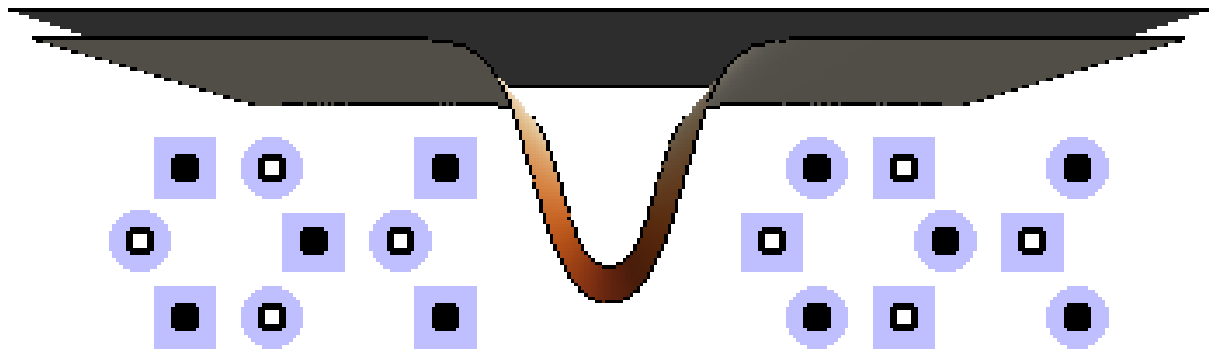}&\includegraphics[scale=.55]{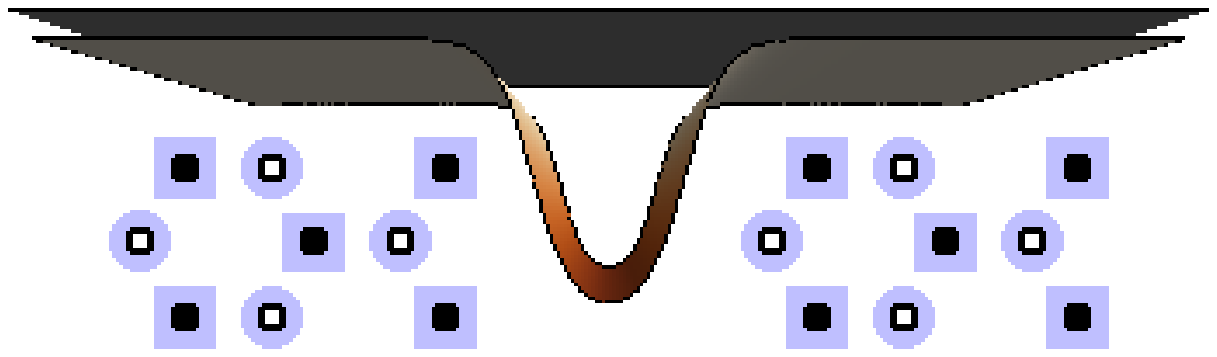}\\
\includegraphics[scale=.6]{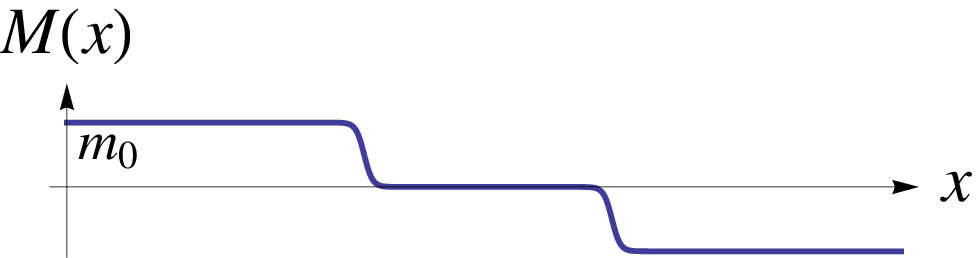}&\includegraphics[scale=.6]{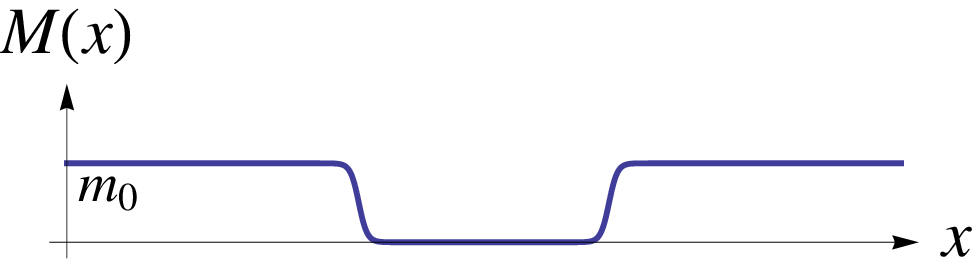}\\
\end{tabular}
\caption{Trench of boron-nitride (brown) covered by graphene sheet (black). Matching of the carbon atoms with boron-nitride on each side of the trench (A-lattice small white dots, B-lattice small black dots, boron blue dots and nitride blue squares) in the inset. The effective masses corresponding to the settings are illustrated below, respectively.\label{figeffmass}}
\label{tab:gt}
\end{figure}

We will consider two situations here. First, the sublattice $A$ is paired with boron on the left side but with nitride on the right side of the trench.  In the second case, the $A$-sublattice is paired with boron on both sides of the trench, see Figure \ref{figeffmass} for illustration.

In the first case when the sublattices are exchanged, the effective mass has different signs on the two sides of the trench. Hence, there are zero-energy bound states in the system. This resembles appearance of zero-modes along the lines with the vanishing effective mass \cite{Semenoff-2008-ID58}.

When the matching of the lattices is the same on both sides of the trench, we have $M(x)\geq 0$ for all $x$.  The effective mass drops from $m_0$ to zero when crossing the trench and then it returns to $m_0$ again. There holds 
\begin{equation}
 W_-=W_+=m_0.
\end{equation}
We suppose that $M(x)=M(-x)$, $M(x)=m_0$ for $|x|>a_1$, $M(x)=0$ for $|x|<a_0$. $M(x)$ it is monotonically increasing from $0$ to $m_0$ for $x\in(a_0,a_1)$. 
We can fit $M(x)^2$ by a piece-wise constant function $M_{T}$ such that
\begin{equation}\label{mupper}
 M(x)^2\leq M_T(x)=\left\{
\begin{array}{ll}m_0^2,& x\in(-\infty,-a_0),
\\ 0,&  x\in [-a_0,a_0],
\\m_0^2,& x\in(a_0,\infty),
\end{array} \right.
\end{equation}
It is convenient to analyze the existence of discrete energies 
with the use of (\ref{cc3}).  Using (\ref{mupper}) and $M_-=m_0$, we get 
\begin{equation}
 \int_{-\infty}^{\infty}(M^2-M_-^2)\leq -2a_0m_0^2
\end{equation}
where the right-hand side is always negative. Hence, we can conclude that there are discrete energy levels in graphene for any finite width of the trench. 

\begin{figure}[h!]
\centering
\includegraphics[scale=1]{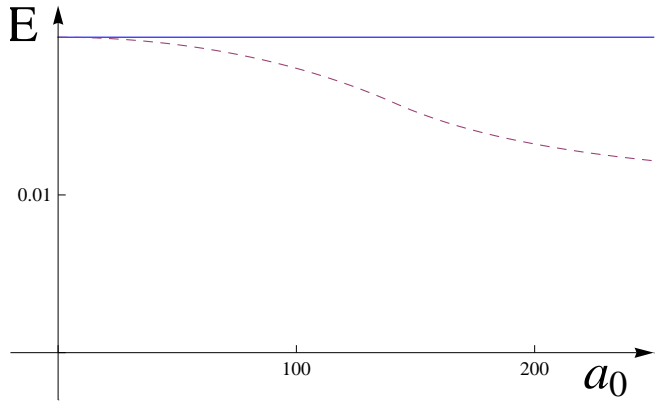}
\caption{The straight line corresponds to the distance $2\sqrt{m_0^2+k_y^2}$ between the thresholds of the continuum of positive and negative energies, the dashed line corresponds to the right-hand side of (\ref{embound}), i.e. to the upper bound of the discrete energy level. We fixed $m=0.01$, $k_y=0$.\label{figgraphgap} }
\end{figure}

Let us make an estimate of the value of the bound state depending on the width of the trench, i.e. in dependence on the parameter $a_0$, 
with the use of (\ref{gapapprox}). We pick up the symmetric test function $\psi_N$ which can be obtained from (\ref{psiN}) by substitution $b=2N$ and $a=N$. The corresponding kinetic term and norm are $\|\psi_N\|^2=\frac{8N}{3}$ and $\|\psi'_N\|^2=\frac{2}{N}$. The term $(\psi_N,M'\psi_N)$ vanishes as the integrand is of odd parity. Let us fix for simplicity $N\geq a_0$, i.e. the test function is constant above the trench. Then we get
\begin{eqnarray}\label{trenchgap}
 (\inf\sigma(h))^2&\leq&\frac{\|\psi'_N\|^2}{\|\psi_N\|^2}
+\frac{(\psi_N,(M^2+k_y^2)\psi_N)}{\|\psi_N\|^2}\leq\frac{\|\psi'_N\|^2}{\|\psi_N\|^2}+\frac{(\psi_N,(M_T^2+k_y^2)\psi_N)}{\|\psi_N\|^2} \\\label{trenchgap2}
  &=&\frac{3}{4N^2}+\left(m_0^2+k_y^2\right)-\frac{3a_0}{4N}m_0^2.
\end{eqnarray}
The formula suggests that the discrete energy separates from the threshold $\sqrt{m_0^2+k_y^2}$ of the essential spectrum as the trench gets wider. It is possible to optimize 
$\frac{3}{4N^2}-\frac{3a_0}{4N}m_0^2$ 
as a function of
$N\in [a_0,\infty)$ 
so 
that the right-hand side would acquire its minimum.  It gets minimized for $N=\max\left\{a_0,\frac{2}{a_0m_0^2}\right\}$ where we take into account the restriction $N\geq a_0$. We get the following estimate for the gap (the upper relation is obtained from (\ref{trenchgap2}) by substitution $N=\frac{2}{a_0m_0^2}$, whereas the lower one by the substitution $N=a_0$)
\begin{equation}\label{embound}
 \Delta(h)\leq\left\{\begin{array}{cl}
2\sqrt{m_0^2+k_y^2-\frac{3a_0^2m_0^4}{16}},& a_0\in\left(0,\frac{\sqrt{2}}{m_0}\right],\\
2\sqrt{\frac{m_0^2}{4}+k_y^2+\frac{3}{4a_0^2}},&  a_0\in\left(\frac{\sqrt{2}}{m_0},\infty\right),                  
                     \end{array}
\right.
\end{equation}
see Figure \ref{figgraphgap}  for illustration.
Notice that the upper bound of the gap converges to $2\sqrt{\frac{m_0^2}{4}+k_y^2}$ for large $a_0$. It indicates that for large values of $a_0$, our requirement $N\geq a_0$ should be revised as the test functions with $N<a_0$ could provide a better estimate of the gap.

\subsection{Circular current loop around a nanotube}
Let us consider the setting where a carbon nanotube forms the axis of a circular current loop. 
The vector potential is given by 
the Biot-Savart law
\begin{equation}\label{vectorpotential}
 \mathbf{A^{mg}}(\mathbf{x})=\frac{\mu_0 I}{4\pi}\oint_{\rm loop}\frac{d\mathbf{x}'}{|\mathbf{x}-\mathbf{x}'|},
\end{equation}
where the line integral is computed along the loop. The vector potential is parallel with the electric current. Hence, it can be written on the surface of the nanotube as $\mathbf{A^{mg}(x)}=\mathbf{e_{\phi}}A_{\phi}(x)$ where  $\mathbf{e_{\phi}}$ is the unit vector tangent to the surface and perpendicular to the axis. In order to compute the explicit form of $\mathbf{A^{mg}}$, let us introduce the following parametrization: in Cartesian coordinates, the points $\mathbf{x'}$ of the loop are given as 
$\mathbf{x}'=R\cos\nu\mathbf{e_z}+R\sin\nu\mathbf{e_y}$ while the points $\mathbf{x}$ on the surface of the nanotube are $\mathbf{x}=r\cos\nu\mathbf{e_z}+r\sin\nu\mathbf{e_y}+x\, \mathbf{e_{x}}$  with $\nu\in[0,2\pi)$. $R$ and $r$ are the radii of the loop and the nanotube, respectively. The system has rotational symmetry which makes it sufficient to evaluate $\mathbf{A^{mg}}(\mathbf{x})$ for fixed point  $\mathbf{x}=r\,\mathbf{e_{z}}+x\mathbf{e_{x}}$ on the nanotube. 

After the substitution $d\mathbf{x}'=R(-\sin\nu\mathbf{e_z}+\cos\nu\mathbf{e_y})d\nu$ in (\ref{vectorpotential}), the vector potential induced by current loop is given by
\begin{equation}\label{Aint}
 \mathbf{A^{mg}}=A_{\phi}\mathbf{e_{y}}=\mathbf{e_{y}}\frac{\mu_0 I}{4\pi}\int_{0}^{2\pi}\frac{\cos\nu d\nu}{\sqrt{1+q^2-2q \cos \nu +v^2}},\quad q=\frac{r}{R},\quad v=\frac{x}{R},
\end{equation}
where the term containing $\mathbf{e_z}\sin\nu$ is canceled out for being the odd function of $\nu$. We passed to the rescaled coordinate $v$.
The integral can be rewritten in terms of the complete elliptic integrals $\mathcal{K}(m)=\int_{0}^{\frac{\pi}{2}}(1-m \sin{\phi}^2)^{-\frac{1}{2}}d\phi$ and $\mathcal{E}(m)=\int_{0}^{\frac{\pi}{2}}\sqrt{1-m \sin{\phi}^2}d\phi$ of the first and the second kind, respectively, in the following manner
\begin{equation}\label{Aelliptic}
 A_{\phi}(x)=\frac{\mu_0 I}{2\pi}\frac{1}{\sqrt{(1+q)^2+v(x)^2}}\left(\frac{(1+q)^2+v(x)^2}{q}\left(\tilde{\mathcal{K}}(x)-\tilde{\mathcal{E}}(x)\right)-2\tilde{\mathcal{K}}(x)\right).
\end{equation} 
We abbreviated here $\tilde{\mathcal{E}}(x)\equiv \mathcal{E}\left(\frac{4q}{(1+q)^2+v(x)^2}\right)$ and $\tilde{\mathcal{K}}(x)\equiv \mathcal{K}\left(\frac{4q}{(1+q)^2+v(x)^2}\right)$. 

Expanding the integrand of  (\ref{Aint}) for large $v$ and integrating term by term over $\nu$, one can find that the vector potential behaves for $|v|\rightarrow+\infty$ as  
\begin{equation}\label{Aasymptotic}
A_{\phi}\sim \frac{\mu_0 I q }{4|v|^3}.
\end{equation} 
Let us notice that the asymptotic behavior is in agreement with the multi-pole expansion of the vector potential around the loop; the dipole term vanishes in our case since the nanotube is coaxial with the loop.

We fix the potential term $W(x)$ of the Hamiltonian (\ref{th}) in the following form 
\begin{equation}\label{Wcircular}
 W(x)=\frac{3a_{cc}e}{2\hbar}A_{\phi}(x)+k_y=\tilde{A_{\phi}}(x)+k_y,\quad  W_{\pm}=k_y,
\end{equation}
where we introduced the notation $\tilde{A_{\phi}}=\frac{3a_{cc}e}{2\hbar}A_{\phi}$ for convenience.  
We suppose that the carbon nanotube is semi-conducting, i.e. $k_y\neq0$ and $M=0$.

The asymptotic behavior of $W^2-W_-^2$  for large $v$ can be found with the use of (\ref{Aasymptotic}), 
\begin{eqnarray}\label{V}
 W^2-W_-^2&=&\tilde{A_{\phi}}^2+2k_y\tilde{A_{\phi}} \sim 2c_1k_y\frac{\pi q }{|v|^3}, \quad c_1=\frac{3a_{cc}e}{2\hbar}\frac{\mu_0 I }{4\pi },
  \end{eqnarray}
It goes to zero from below or above, depending on the sign of $c_1k_y$. From now on, we will fix $c_1k_y<0$ with $c_1>0$ (i.e. $I>0$) and $k_y<0$. 

The essential spectrum of the Hamiltonian is $\sigma_{\rm ess}(h)=\left(-\infty,-|k_y|\right]\cup\left[\,|k_y|,\infty\right)$. There are no zero modes in the system since $W_+=W_-$ (see No. 3 of the theorem).  To test the system on presence of non-zero discrete energies,
it is convenient to use the condition (\ref{cc3}). It reads 
\begin{equation}\label{qs}
 \int_{-\infty}^{\infty}\left(\tilde{A}_{\phi}^2(x)-2|k_y|\tilde{A}_{\phi}(x)\right)dx<0.
\end{equation}
The integral is too complicated to be computed explicitly. We shall find the upper bound of the integrand that would be easier to integrate. We shall find it in the following form 
\begin{equation}\label{Upperbound}
(\tilde{A}_{\phi}^{\max})^2-2|k_y|\tilde{A}_{\phi}^{\min}\geq \tilde{A}_{\phi}^2-2|k_y|\tilde{A}_{\phi}
\end{equation}
where $\tilde{A}_{\phi}^{\max}(v)\geq \tilde{A}_{\phi}(v)\geq \tilde{A}_{\phi}^{\min}(v)$ for all $v$. 

It is convenient to pick up the definition of the complete elliptic integrals in terms of the infinite series \cite{abramovitz}. 
For $|z|<1$, there hold the following formulas 
\begin{equation}\label{KEpower}
 \mathcal{E}(z)=\frac{\pi}{2}\sum_{n=0}^{\infty}\frac{\left(-\frac{1}{2}\right)_n\left(\frac{1}{2}\right)_n}{n!^2}z^n,\quad 
\mathcal{K}(z)=\frac{\pi}{2}\sum_{n=0}^{\infty}\frac{\left(\frac{1}{2}\right)_n\left(\frac{1}{2}\right)_n}{n!^2}z^n,
\end{equation}
where $ \left(-\frac{1}{2}\right)_n=-\frac{(2n-2)!}{2^{2n-1}(n-1)!},$ $ \left(\frac{1}{2}\right)_n=\frac{(2n-1)!}{2^{2n-1}(n-1)!}$ are Pochhammer symbols. Using them, we can find (see Appendix~\ref{App2} for details)
\begin{eqnarray}\label{Amaxmin}
 {A}^{\max}_{\phi}(v)&\equiv&\frac{c_1}{\sqrt{(1+q)^2+v^2}}
\left(
2\mathcal{K}^{\rm max}\left(\frac{4q}{(1+q)^2+v^2}\right)-\sum_{n=0}^{2}\frac{1}{n+1}\frac{\left(\frac{1}{2}\right)^2_n\left(\frac{4q}{(1+q)^2+v^2}\right)^n}{(n!)^2}
\right),\\
A_{\phi}^{\min}(v)&\equiv& \frac{c_1}{\sqrt{(1+q)^2+v^2}}
\left(\pi\sum_{n=0}^{2}\frac{n}{n+1}\frac{\left(\frac{1}{2}\right)^2_n\left(\frac{4q}{(1+q)^2+v^2}\right)^n}{n!^2}\right),
\end{eqnarray}
where
\begin{eqnarray}\label{upperK}
 \mathcal{K}^{\rm max}(z)&\equiv&\frac{\pi}{2\sqrt{1-z}}+\sum_{n=1}^{2}\left(\frac{\left(\frac{1}{2}\right)^2_nz^n}{n!^2}-\left(\frac{\pi}{2\sqrt{1-z}}\right)^{(n)}_{|z=0}\frac{z^n}{n!}\right)\geq\mathcal{K}(z).
\end{eqnarray}

Substituting in the left-hand side of (\ref{Upperbound}), 
it can be integrated analytically. We get 
\begin{equation}\label{intcond}
 \int_{-\infty}^{\infty}((A^{\max}_{\phi})^2-2|k_y| A^{\min}_{\phi})=F(q,k_y,c_1),
\end{equation}
where
\begin{eqnarray}\label{F}
 F(q,k_y,c_1)&=&-\frac{c_1 |k_y| \pi  q \left(2+7 q+2 q^2\right)}{(1+q)^4}+c_1^2 \pi ^2 \left(\frac{\pi  \left(\frac{32}{1-q}+\frac{315 q^4}{128 (1+q)^8}+\frac{45 q^3}{4 (1+q)^6}+\frac{45 q^2}{2 (1+q)^4}+\frac{24 q}{(1+q)^2}\right)}{16 (1+q)}\right.\nonumber\\&&\left.-\frac{4 \mathcal{K}\left(-\frac{4 q}{(1-q)^2}\right)}{1-q}+\frac{2 q \mathcal{E}\left(\frac{4 q}{(1+q)^2}\right)-(13+q (28+13 q)) \left(\mathcal{K}\left(\frac{4 q}{(1+q)^2}\right)-\mathcal{E}\left(\frac{4 q}{(1+q)^2}\right)\right)}{8 (1+q)^3}\right).
\end{eqnarray}
When $F(q,k_y,c_1)<0$, there are discrete energy levels in the system. 

To assess the sign of $F(q,k_y,c_1)$, we have to specify the physically reasonable range of the parameters $c_1$, $q$ and $k_y$.
We suppose the current loop to be made of another nanotube. The currents supported by the carbon nanotubes can go up to $I\approx 25\mu A$, see \cite{McEuen-2002-ID448}. Therefore, the parameter $c_1$  is quite small. As $c_1\approx 0.032 A^{-1} I $, we take $c_1\approx 10^{-7}$. For the ratio $q$ of the radius of the nanotube and the current loop, we find the values as in the range $q\approx 10^{-4}$ to $q\approx 10^{-2}$ to be experimentally feasible. We take the radius of the nanotube $r\approx 13$ in the units introduced in the text above (\ref{vp}). Hence, we have $k_y=-\frac{1}{3r}\sim -10^{-2}$.
For the considered small values of $c_1$ and for $q\gg c_1$, we can see that the first negative term in (\ref{F}) becomes dominant as the rest of the expression depends on $c_1^2$. The integral (\ref{intcond}) gets negative, 
implying the existence of bound states in the spectrum.

\begin{figure}[h!]
\centering
\includegraphics{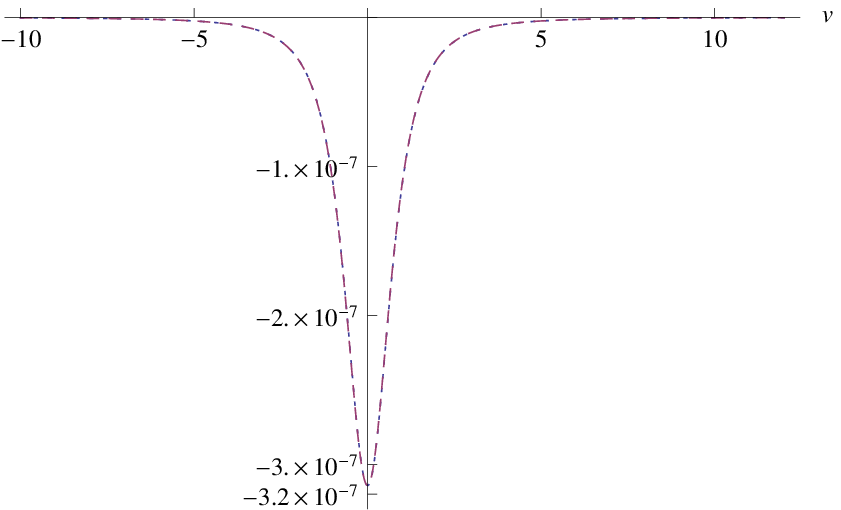}
\includegraphics[scale=.91]{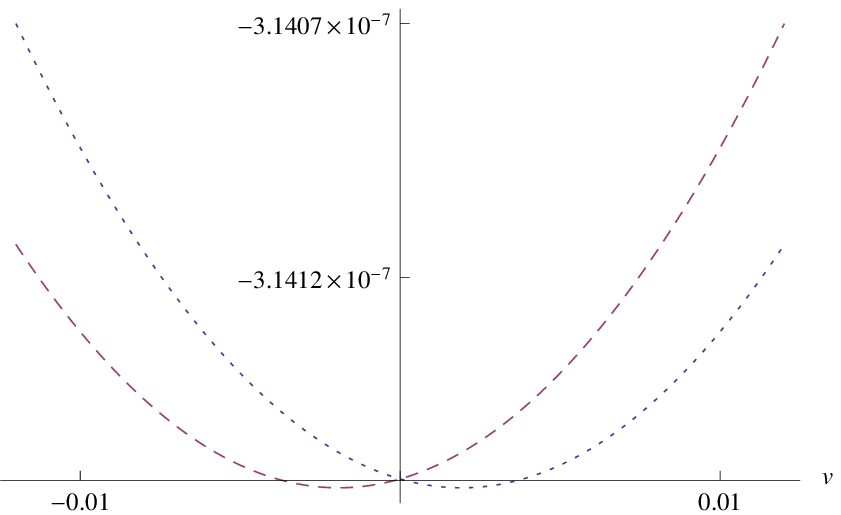}
\caption{\label{V00}Left: plot of $\frac{\tilde{A}_{\phi}^2-2|k_y|\tilde{A}_{\phi}\pm\tilde{A}'_{\phi}}{k_y^2}$. The two functions are almost identical. Right: The magnified sector around the minimum of the two functions. In both figures, we fixed $k_y=-2\times 10^{-2}$, $c_1=5\times10^{-7}$, $q=2\times 10^{-3}$ and $R=1.3\times 10^4$ in the units introduced in the text above (\ref{vp}).}
\end{figure}

The essential spectrum of the system is 
\begin{equation}\sigma_{\rm ess}(h)=\left(-\infty,-|k_y|\right]\cup\left[|k_y|,\infty\right).
\end{equation}
When $F(q,k_y,c_1)<0$, there are discrete energy levels $|\lambda_{\rm disc}|<|k_y|$. However, for the considered range of physical parameters, the distance of the discrete energy $\lambda_{\rm disc}$ from the threshold $|k_y|$ of the essential spectrum is very small.  We can use the formula (\ref{V0}) for estimating the gap. The discrete energy level lies above $\sqrt{V_0}$, 
where $V_0=\min\left\{\inf_{x\in\mathbb{R},\epsilon\in\{-1,1\}} (\tilde{A}_{\phi}^2(x)-2|k_y|\tilde{A}_{\phi}(x)+\epsilon\tilde{A}'_{\phi}(x))+k_y^2\right\}$. 
Instead of further investigation of the exact values of $V_0$, let us make the estimate of its value by graphical analysis. 
In Figure \ref{V00}, there is a plot of dimensionless functions $\frac{\tilde{A}_{\phi}^2-2|k_y|\tilde{A}_{\phi}\pm\tilde{A}'_{\phi}}{k_y^2}$. The minimum of the functions coincides with $\frac{V_0-k_y^2}{k_y^2}$. For the given fixed constants, we get $\frac{V_0-k_y^2}{k_y^2}\in (-3.2\times 10^{-7},-3.14\times 10^{-7})$. The discrete energy level $\lambda_{\rm disc}$ then satisfies
\begin{equation}
 0<\frac{k_y^2-\lambda_{\rm disc}^2}{k_y^2} < 3.2\times 10^{-7}.
\end{equation}

\section{Discussion and Outlook}
In the paper, we considered confinement of Dirac fermions in graphene and carbon nanotubes by external magnetic field, mechanical deformations and by inhomogeneities in the substrate. We focused on the settings with translational invariance in one direction and provided a set of conditions that are sufficient for confinement. The theoretical analysis was performed with the use of the variational principle that was applied on the square of the Dirac Hamiltonian. The results were summarized in the form of theorem in section \ref{Theorem} and illustrated in the examples of realistic systems. 

The sufficient conditions (integral inequalities) 
were derived under general assumptions where only the asymptotic properties of the interaction were anticipated. 
When more precise properties of the interaction are known, the sufficient conditions can be further precised to offer finer test on presence of discrete energies. We believe that the step-by-step presentation provided by the current paper will allow one to follow the procedure in these cases. 

The analysis of the discrete energies is important for understanding of the optical properties of graphene and related carbon nanostructures. Interaction of Dirac electrons with the electromagnetic radiation was discussed in the context of the Dirac equation mostly for free graphene \cite{Nair-2008-ID443}, \cite{Calvo-2011-ID408}, \cite{Farias-2013-ID446}. We believe that the current results can be useful for the analysis of the optical properties (optical absorption in particular \cite{Berciaud-2007-ID405}), of the systems that are subject to the external fields or mechanical deformations.

Our qualitative results can provide further insight into the existing explicit models discussed in the literature.  They can be used directly for the  analysis of the linear barriers in graphene or the trenches in the substrate \cite{Pereira-2009-ID199}, \cite{Neek-Amal-2012-ID438}. They can be also relevant for the systems where space-dependent chemical potential appears in the graphene-based heterostructures \cite{halterman}.
By the presented rigorous analysis, the work also contributes to the series of the recent papers where the spectral properties of Dirac operators were discussed \cite{Petr}, \cite{2013arXiv1303.2185E}, \cite{Downes-2013-ID451}, \cite{Pankrashkin-2014-ID398}, \cite{Aiba-2014-ID400}, or \cite{Morozov-2014-ID401}. In particular, let us mention \cite{Kandemir-2013-ID43} where the variational principle was utilized for analysis of the bound states in the quantum anti-dots. 

We explained how the powerful variational approach
can be used to study bound states of Dirac systems,
even if it cannot be applied directly.
It would be interesting to consider planar systems where the Dirac Hamiltonian ceases to be separable due to the lack of symmetries. These systems are realized experimentally, e.g. in the form of graphene quantum dots or heterostructures with structured substrate.  The Dirac Hamiltonian describing particles in the curved geometry appears in description of Dirac fermions in graphene with ripples and bumps \cite{deJuan-2007-ID286} or in description of graphene based macromolecules, e.g. fullerenes \cite{Gonzalez-1992-ID294}. Extension of our current results on this kind of systems would be definitely very interesting. In this context, it is worth noticing that Dirac fermions in graphene can be utilized as an interesting model of quantum field theory in the curved space \cite{Iorio-2012-ID163}, \cite{Iorio-2013-ID434}.

In the work we focused on the systems where either the vector potential or the effective mass was position dependent, but not both at the same time. The electrostatic potential was assumed to be absent in all the considered systems. This allowed us to diagonalize the Hamiltonian into the form of a Schr\"odinger operator. It would be interesting to consider a wider class of systems, where both vector potential and effective mass are inhomogeneous and/or the electrostatic potential is present. The variational approach could generalize the existing results \cite{dot}, where the magnetic and electrostatic fields with specific asymptotics were considered and the necessary condition for existence of bound states was discussed.  Detailed discussion of all these problems, attractive from both mathematical and physical point of view, should be aimed in the future.

\appendix

\section{Location of the essential spectrum}\label{App1}
We shall prove (\ref{essential}).
First, let us verify that  $\sigma_\mathrm{ess}(H)\subset[V_-,\infty)$. It is convenient to find a new operator $H^R$ which has,  when compared to $H$, a lower threshold of the essential spectrum, 
\begin{equation}\label{ess}
\inf\sigma_{\rm ess}(H)\geq\inf \sigma_{\rm ess}(H^R),                                                                                                                                  
\end{equation}
and, at the same time, $\inf \sigma_{\rm ess}(H^R)$ can be bounded from below by a constant that goes to $V_-$ for large $R$. 

Let us fix $H^R$ as a direct sum of three operators, 
$H^R=H_{(-\infty,-R)}\oplus H_{(-R,R)}\oplus H_{(R,\infty)}$. Here, the operator $H_{I}$ acts as $H$, however,
its domain is different; it is defined on functions that are integrable up to their second derivative on $I$ and satisfy Neumann boundary conditions 
at $\pm R$. Then we have $H\geq H^R$ which follows from the fact that $D(q_{H})\subset D(q_{H^R})$ (the functions from $D(q_{H})$ have to be continuous, whereas the functions from $D(q_{H^R})$ can have discontinuities at $\pm R$). Together with the minimax principle (see ch. 13.1 in \cite{R-S}), it implies validity of (\ref{ess}). 
 
The operator $H_{(-R,R)}$ is defined 
on the compact interval with Neumann boundary conditions. As it has purely discrete spectrum, it has no effect on the threshold of the essential spectrum of $H^R$. We have
\begin{equation}\label{essN}
 \inf \sigma_{\rm ess}(H^R)=\min\{\inf \sigma_{\rm ess}(H_{(-\infty,-R)}),\inf \sigma_{\rm ess}(H_{(R,\infty)})\}
\end{equation}
Here we can see the advantage of the Neumann bracketing; it allows to simplify the analysis of the essential spectrum by 
considering the asymptotic regions of the potential only, where its behavior is controlled 
by (\ref{prop}). Fixing $\epsilon>0$, we can set $R$ such that $|V(x)-V_{\pm}|<\epsilon$ 
for all $|x|>R$. Employing the variational principle (\ref{variational})
and  the fact that $\inf \sigma_{\rm ess}\left(H_{(-\infty,-R)}\right)\geq\inf \sigma\left(H_{(-\infty,-R)}\right)$, we can write
\begin{equation}
 \inf \sigma_{\rm ess}\left(H_{(-\infty,-R)}\right)\geq\inf \sigma\left(H_{(-\infty,-R)}\right)\geq V_--\epsilon. 
\end{equation}
We can find in the same vein that $\inf \sigma_{\rm ess}\left(H_{(R,\infty)}\right)\geq V_+-\epsilon$. Since $\epsilon$ is an arbitrary positive number and there holds $V_-\leq V_+$, (\ref{ess}) and (\ref{essN}), we have $\sigma_{\rm ess}(H)\subset [V_-,\infty)$.

Now, let us show that there also holds $\sigma_{\rm ess}(H)\supset [V_-,\infty).$ It is sufficient to show 
that for any $\lambda \geq V_-$, there exists a normalized sequence of functions $\psi_n$ from the domain 
of $H$ such that $(H-\lambda)\psi_n{\rightarrow}0$ for ${n\rightarrow\infty}$. Then (by Weyl's 
criterion \cite{R-S}), $\lambda$ belongs to the spectrum of $H$. We can define 
$\psi_n=\frac{\phi\left(\frac{x}{n}+n\right)}{\sqrt{n}}e^{ikx}$ where $k\in\mathbb{R}$, $k^2=\lambda$ and $\phi(x)$ 
is a smooth normalized real-valued function that is nonzero on the interval $(-1,1)$. Hence, $\psi_{n}$ is nonzero 
on the interval $I_n=(-n^2-n,-n^2+n)$. Direct computation then yields
\begin{equation}
 \|(H-k^2-V_-)\psi_n\|\leq\frac{\|\phi''\|}{n^2}+2|k|\frac{\|\phi'\|}{n}+\sup_{I_n}{|V-V_-|}.
\end{equation}
The right-hand side tends to zero as $n\rightarrow \infty$. It means that any $\lambda=k^2+V_-\geq V_-$ is 
in the spectrum of $H$, which completes the proof of (\ref{essential}).

We claim below (\ref{hPsi}) that the formula can be used to show that there holds 
\begin{equation}
 \lambda\in\sigma(h)\Rightarrow -\lambda\in\sigma(h),\quad \lambda^2\neq M^2.
\end{equation}
Let $\lambda^2\in\sigma(H)$. By Weyl's criterion, there is a sequence of normalized states $\psi_n$ from the domain of $H$ such that 
$(H-\lambda^2)\psi_n{\rightarrow}0$ for ${n\rightarrow\infty}$. Let us construct the sequence of spinors 
\begin{equation}
    \Psi^{\pm }_n =\alpha_n\left(\begin{array}{c}\frac{{-i(\partial_x+W(x))}}{\pm |\lambda|-M}\psi_n\\\psi_n\end{array}\right),                                                                                 \end{equation}
where $\alpha_n$ is fixed by requirement $\|\Psi_n\|=1$. One can check that $\alpha_n$ is bounded for all $n$ as long as $\lambda\neq 0$. Then we find that
\begin{equation}
 \|(h\mp|\lambda|)\Psi^{\pm}_n\|=\alpha_N\left(\begin{array}{c}0\\\frac{H-|\lambda|^2}{-M\pm |\lambda|}\psi_n\end{array}\right)
\end{equation}
which tends to $0$ for $N\rightarrow \infty$. As $\Psi_N^{\pm}$ is normalized function from the domain of $h$, then Weyl's criterion implies that both $|\lambda|$ and $-|\lambda|$ are in the spectrum of $h$. 

\section{Approximation of the vector potential of the current loop}\label{App2}
For reals $z$ where $|z|<1$, there hold the following formulas 
\begin{equation}
 \mathcal{E}(z)=\frac{\pi}{2}\sum_{n=0}^{\infty}\frac{\left(-\frac{1}{2}\right)_n\left(\frac{1}{2}\right)_n}{n!^2}z^n,\quad 
\mathcal{K}(z)=\frac{\pi}{2}\sum_{n=0}^{\infty}\frac{\left(\frac{1}{2}\right)_n\left(\frac{1}{2}\right)_n}{n!^2}z^n,
\end{equation}
where $ \left(-\frac{1}{2}\right)_n=-\frac{(2n-2)!}{2^{2n-1}(n-1)!},$ $ \left(\frac{1}{2}\right)_n=\frac{(2n-1)!}{2^{2n-1}(n-1)!}$ are Pochhammer symbols.

We shall find the upper estimate of $A_{\phi}^2-2|k_y|A_{\phi}$, see  (\ref{Upperbound}),
\begin{equation}
(A^{\max}_{\phi})^2-2|k_y| A^{\min}_{\phi}\geq A_{\phi}^2-2|k_y|A_{\phi}
\end{equation}
where 
\begin{equation}
 A_{\phi}=\frac{c_1}{\sqrt{(1+q)^2+v^2}}\left(\frac{(1+q)^2+v^2}{q}\left(\mathcal{K}-\mathcal{E}\right)-2\mathcal{K}\right),\quad q=\frac{r}{R},\quad v=\frac{x}{R}
\end{equation}
and  $\mathcal{K}\equiv \mathcal{K}\left(\frac{4q}{(1+q)^2+v^2}\right)$, $\mathcal{E}\equiv \mathcal{E}\left(\frac{4q}{(1+q)^2+v^2}\right)$.

First, it is convenient to find the upper bound $\mathcal{K}^{\rm max}(z)$ of $\mathcal{K}(z)$,
\begin{eqnarray}
 \mathcal{K}^{\rm max}(z)&\geq& \mathcal{K}(z),\\
 \mathcal{K}^{\rm max}(z)&=&\frac{\pi}{2\sqrt{1-z}}+\sum_{n=1}^{N_0}\left(\frac{\left(\frac{1}{2}\right)^2_nz^n}{n!^2}-\left(\frac{\pi}{2\sqrt{1-z}}\right)'_{z=0}\frac{z^n}{n!}\right).
\end{eqnarray}
Here, the first term is obtained from $\int_{0}^{\frac{\pi}{2}}(1-z \sin{\phi}^2)^{-\frac{1}{2}}d\phi$ by the substitution $\sin{\phi}^2\rightarrow 1$. As it is not accurate enough, we fix the first $N_0$ coefficients in its series expansion such that they coincide with with those of $\mathcal{K}(z)$. 

Then we focus on the following function which forms the ``core'' of $A_{\phi}$. Using the expansion series for the complete elliptic integrals, we can write
\begin{eqnarray}\label{core}
 \frac{4}{z}(\mathcal{K}(z)-\mathcal{E}(z))-2\mathcal{K}(z)&=&\frac{\pi}{2}\sum_{n=1}^{\infty}\frac{\left(\frac{1}{2}\right)^2_n\left(\frac{2n}{2n-1}\right)z^n}{n!^2}-\frac{\pi}{2}\sum_{n=0}\frac{\left(\frac{1}{2}\right)_n^2z^n}{n!^2}\nonumber\\
 &=&\pi\sum_{n=0}^{\infty}\frac{n}{n+1}\frac{\left(\frac{1}{2}\right)^2_nz^n}{n!^2}=2\mathcal{K}(z)-\pi\sum_{n=0}^{\infty}\frac{1}{n+1}\frac{\left(\frac{1}{2}\right)^2_nz^n}{n!^2}.
\end{eqnarray}
We can find the lower and the upper bound of the left-hand side
\begin{eqnarray}
 \frac{4}{z}(\mathcal{K}(z)-\mathcal{E}(z))-2\mathcal{K}(z)&\geq&\pi\sum_{n=0}^{N_1}\frac{n}{n+1}\frac{\left(\frac{1}{2}\right)^2_nz^n}{(n!)^2},
\quad N_1\in\mathbb{N},\\
\frac{4}{z}(\mathcal{K}(z)-\mathcal{E}(z))-2\mathcal{K}(z)&\leq& 2\mathcal{K}(z)-\pi\sum_{n=0}^{N_2}\frac{1}{n+1}\frac{\left(\frac{1}{2}\right)^2_nz^n}{(n!)^2}\nonumber\\
&\leq& 2\mathcal{K}^{\rm max}(z)-\pi\sum_{n=0}^{N_2}\frac{1}{n+1}\frac{\left(\frac{1}{2}\right)^2_nz^n}{(n!)^2},\quad N_2\in\mathbb{N}.
\end{eqnarray}

Now, we can introduce the upper and the lower bound 
$A_{\phi}^\mathrm{max}$ and $A_{\phi}^\mathrm{min}$ 
that satisfy $A_{\phi}^{\rm max}\geq A_{\phi}\geq A_{\phi}^\mathrm{min}$,
\begin{eqnarray}
 \mathcal{A}^\mathrm{max}_{\phi}&=&\frac{\mu_0I}{2\pi}\frac{1}{\sqrt{(1+q)^2+v^2}}
\left(
2\mathcal{K}^{\rm max}\left(\frac{4q}{(1+q)^2+v^2}\right)-\sum_{n=0}^{N_2}\frac{1}{n+1}\frac{\left(\frac{1}{2}\right)^2_n\left(\frac{4q}{(1+q)^2+v^2}\right)^n}{(n!)^2}
\right),\\
 A^\mathrm{min}_{\phi}&=&\frac{c_1}{\sqrt{(1+q)^2+v^2}}
\left(\pi\sum_{n=0}^{N_4}\frac{n}{n+1}\frac{\left(\frac{1}{2}\right)^2_nz^n}{n!^2}\right).
\end{eqnarray}

\section*{Acknowledgements}
V.J. would like to thank prof.~Francisco Fernandez for discussion.
The research was partially supported 
by RVO61389005 and the GACR grant No.\ 14-06818S.



\begin{thebibliography}{10}

\bibitem{Semenoff-1984-ID345}
G.~W. Semenoff,
\newblock Physical Review Letters {\bf 53}, 2449 (1984).

\bibitem{Novoselov-2005-ID389}
K.~S. Novoselov {\em et~al.},
\newblock Nature {\bf 438}, 197 (2005).

\bibitem{Zhang-2005-ID435}
Y.~Zhang, Y.-W. Tan, H.~L. Stormer, and P.~Kim,
\newblock Nature {\bf 438}, 201 (2005).

\bibitem{Nair-2008-ID443}
R.~R. Nair {\em et~al.},
\newblock Science {\bf 320}, 1308 (2008).

\bibitem{Katsnelson-2006-ID444}
M.~I. Katsnelson, K.~S. Novoselov, and A.~K. Geim,
\newblock Nature Physics {\bf 2}, 620 (2006).

\bibitem{Ando-1998-ID308}
T.~Ando, T.~Nakanishi, and R.~Saito, 
\newblock J Phys Soc Jap {\bf 67}, 2857 (1998).

\bibitem{Jakubsky-2011-ID20}
V.~Jakubsk{\'y}, L.-M. Nieto, and M.~S. Plyushchay,
\newblock Physical Review D {\bf 83}, 47702 (2011).

\bibitem{Downing-2011-ID450}
C.~A. Downing, D.~A. Stone, and M.~E. Portnoi,
\newblock Physical Review B {\bf 84}, 155437 (2011).

\bibitem{Downing2} D.~A.~Stone, C.~A.~Downing, and M.~E.~Portnoi, Physical Review B {\bf 86}, 075464 (2012).

\bibitem{Pereira-2006-ID313}
J.~M. Pereira, V.~Mlinar, F.~M. Peeters, and P.~Vasilopoulos,
\newblock Physical Review B {\bf 74}, 45424 (2006).

\bibitem{Hartmann-2010-ID45}
R.~R. Hartmann, N.~J. Robinson, and M.~E. Portnoi,
\newblock Physical Review B {\bf 81}, 245431 (2010).

\bibitem{Hartmann-2013-ID413}
R.~R. Hartmann and M.~E. Portnoi,
\newblock Physical Review A {\bf89}, 012101 (2014).

\bibitem{review}A.~V.~Rozhkov, G.~Giavaras, Yury~P.~Bliokh, V.~Freilikher, and F.~Nori, Physics Reports {\bf 503}, 77 (2011).

\bibitem{Peres-2006-ID433}
N.~M.~R. Peres, A.~H. {Castro Neto}, and F.~Guinea,
\newblock Physical Review B {\bf 73}, 241403 (2006).

\bibitem{deMartino-2007-ID53}
A.~de~Martino, L.~Dell'Anna, and R.~Egger,
\newblock Physical Review Letters {\bf 98}, 66802 (2007).

\bibitem{RamezaniMasir-2008-ID440}
M.~{Ramezani Masir}, P.~Vasilopoulos, A.~Matulis, and F.~M. Peeters,
\newblock Physical Review B {\bf 77}, 235443 (2008).

\bibitem{Dell'Anna-2009-ID12}
L.~Dell'Anna and A.~de~Martino,
\newblock Physical Review B {\bf 79}, 45420 (2009).

\bibitem{RamezaniMasir-2011-ID3}
M.~{Ramezani Masir}, P.~Vasilopoulos, and F.~M. Peeters,
\newblock J Phys Condens Matter {\bf 23}, 315301 (2011).

\bibitem{RamezaniMasir-2009-ID95}
M.~{Ramezani Masir}, P.~Vasilopoulos, and F.~M. Peeters,
\newblock New Journal of Physics {\bf 11}, 095009 (2009).

\bibitem{Roy-2012-ID415}
P.~Roy, T.~{Kanti Ghosh}, and K.~Bhattacharya,
\newblock J Phys Condens Matter {\bf 24}, 5301 (2012).

\bibitem{tarun} T.~{Kanti Ghosh}, J Phys Condens Matter {\bf 21}, 045505 (2009).

\bibitem{Kuru-2009-ID442}
S.~Kuru, J.~Negro, and L.~M. Nieto,
\newblock J Phys Condens Matter {\bf 21}, 455305 (2009).

\bibitem{Milpas-2011-ID92}
E.~Milpas, M.~Torres, and G.~Murgu{\'i}a,
\newblock J Phys Condens Matter {\bf 23}, 245304 (2011).

\bibitem{Midya-2014-ID429}
B.~Midya and D.~J. Fern{\'a}ndez,
\newblock arxiv: 1402.4584v1  (2014).

\bibitem{Jakubsky-2012-ID85}
V.~Jakubsk{\'y} and M.~S. Plyushchay,
\newblock Physical Review D {\bf 85}, 045035 (2012).

\bibitem{Myoung-2011-ID54}
N.~Myoung, G.~Ihm, and S.~J. Lee,
\newblock Physical Review B {\bf 83}, 113407 (2011).

\bibitem{Kane-1997-ID103}
C.~L. Kane and E.~J. Mele,
\newblock Physical Review Letters {\bf 78}, 1932 (1997).

\bibitem{Suzuura-2002-ID104}
H.~Suzuura and T.~Ando,
\newblock Physical Review B {\bf 65}, 235412 (2002).

\bibitem{Vozmediano-2010-ID97}
M.~A.~H. Vozmediano, M.~I. Katsnelson, and F.~Guinea,
\newblock Physics Reports {\bf 496}, 109 (2010).

\bibitem{Pereira-2009-ID199}
V.~M. Pereira and A.~H. {Castro Neto},
\newblock Physical Review Letters {\bf 103}, 046801 (2009).



\bibitem{Semenoff-2008-ID58}
G.~W. Semenoff, V.~Semenoff, and F.~Zhou,
\newblock Physical Review Letters {\bf 101}, 87204 (2008).

\bibitem{mass1} G.~Giavaras and F.~Nori, Applied Physics Letters {\bf97}, 243106 (2010).

\bibitem{mass2} G.~Giavaras and F.~Nori, Physics Review B {\bf83}, 165427 (2011).

\bibitem{mass3} C.~Popovici, O.~Oliveira, W.~de Paula, and T.~Frederico, Physics Review B {\bf 85}, 235424 (2012). 



\bibitem{Katsnelson}
M.~I. Katsnelson,
\newblock {\em Graphene: Carbon in Two Dimensions} (Cambridge University Press,
  2012).

\bibitem{Correa-2013-ID86}
F.~Correa and V.~Jakubsk{\'y},
\newblock Physical Review D {\bf 87}, 085019 (2013).

\bibitem{vanRoy-1993-ID2}
W.~van Roy {\em et~al.},
\newblock Journal of Magnetism and Magnetic Materials {\bf 121}, 197 (1993).

\bibitem{Charlier-2007-ID169}
J.-C. Charlier, X.~Blase, and S.~Roche,
\newblock Reviews of Modern Physics {\bf 79}, 677 (2007).

\bibitem{Clauss-1998-ID436}
W.~Clauss, D.~J. Bergeron, and A.~T. Johnson,
\newblock Physical Review B {\bf 58}, 4266 (1998).

\bibitem{Meyer-2005-ID437}
J.~C. Meyer, M.~Paillet, and S.~Roth,
\newblock Science {\bf 309}, 1539 (2005).

\bibitem{Peng-2007-ID186}
H.~B. Peng, C.~W. Chang, S.~Aloni, T.~D. Yuzvinsky, and A.~Zettl,
\newblock Physical Review B {\bf 76}, 35405 (2007).

\bibitem{Fennimore-2003-ID90}
A.~M. Fennimore {\em et~al.},
\newblock Nature {\bf 424}, 408 (2003).

\bibitem{Joselevich-2006-ID184}
E.~Joselevich,
\newblock Chemphyschem {\bf 7}, 1405 (2006).

\bibitem{Giovannetti-2007-ID6}
G.~Giovannetti, P.~A. Khomyakov, G.~Brocks, P.~J. Kelly, and J.~van~den Brink,
\newblock Physical Review B {\bf 76}, 73103 (2007).

\bibitem{Yankowitz-2012-ID4}
M.~Yankowitz {\em et~al.},
\newblock Nature Physics {\bf 8}, 382 (2012).

\bibitem{Sachs-2011-ID7}
B.~Sachs, T.~O. Wehling, M.~I. Katsnelson, and A.~I. Lichtenstein,
\newblock Physical Review B {\bf 84}, 195414 (2011).

\bibitem{Song-2013-ID9}
J.~Song, A.~Shytov, and L.~Levitov,
\newblock Physical Review Letters {\bf 111}, 266801 (2013).

\bibitem{Hunt-2013-ID10}
B.~Hunt {\em et~al.},
\newblock Science {\bf 340}, 1427 (2013).

\bibitem{R-S}

\newblock M.~Reed and B.~Simon, {\em Methods of Modern Mathematical Physics}
  Vol.~4 (Academic Press, 1978).

\bibitem{abramovitz}
M.~Abramowitz and I.~A. Stegun,
\newblock {\em Handbook of Mathematical Functions: with Formulas, Graphs, and
  Mathematical Tables} (Dover Publications, 1965).

\bibitem{McEuen-2002-ID448}
P.~L. McEuen, M.~S. Fuhrer, and H.~Park,
\newblock IEEE Transactions On Nanotechnology {\bf 1}, 78 (2002).


\bibitem{Farias-2013-ID446}
M.~B. Far{\'i}as, G.~F. Quinteiro, and P.~I. Tamborenea,
\newblock The European Physical Journal B {\bf 86}, 432 (2013).

\bibitem{Calvo-2011-ID408}
H.~L. Calvo, H.~M. Pastawski, S.~Roche, and L.~E. F.~F. Torres,
\newblock Applied Physics Letters {\bf 98}, 2103 (2011).

\bibitem{Berciaud-2007-ID405}
S.~Berciaud, L.~Cognet, P.~Poulin, R.~B. Weisman, and B.~Lounis,
\newblock Nano Letters {\bf 7}, 1203 (2007).



\bibitem{Petr}
P.~Freitas and P.~Siegl, preprint
\newblock (2013).

\bibitem{2013arXiv1303.2185E}
D.~M. {Elton}, M.~{Levitin}, and I.~{Polterovich},
\newblock Annales Henri Poincare, DOI:10.1007/s00023-013-0304-2


\bibitem{Downes-2013-ID451}
R.~J. Downes, M.~Levitin, and D.~Vassiliev,
\newblock Journal of Mathematical Physics {\bf 54}, 1503 (2013).

\bibitem{Pankrashkin-2014-ID398}
K.~Pankrashkin and S.~Richard, 
\newblock J.\ Math.\ Phys.\ {\bf55}, 062305 (2014).

\bibitem{Aiba-2014-ID400}
D.~Aiba,
\newblock arxiv:1401.8043v1  (2014).

\bibitem{Morozov-2014-ID401}
S.~Morozov and D.~M{\"u}ller,
\newblock arxiv:1401.5916v1  (2014).

\bibitem{Kandemir-2013-ID43}
B.~S. Kandemir and G.~Omer,
\newblock The European Physical Journal B {\bf 86}, 299 (2013).

\bibitem{Neek-Amal-2012-ID438}
M.~Neek-Amal and F.~M. Peeters,
\newblock Physical Review B {\bf 85}, 195445 (2012).

\bibitem{halterman}K.~Halterman, O.~T.~Valls, M.~Alidoust, Phys.\ Rev.\ Lett.\ {\bf111}, 046602 (2013).

\bibitem{deJuan-2007-ID286}
F.~de~Juan, A.~Cortijo, and M.~A.~H. Vozmediano,
\newblock Physical Review B {\bf 76}, 165409 (2007).

\bibitem{Gonzalez-1992-ID294}
J.~Gonz{\'a}lez, F.~Guinea, and M.~A.~H. Vozmediano,
\newblock Physical Review Letters {\bf 69}, 172 (1992).

\bibitem{Iorio-2012-ID163}
A.~Iorio and G.~Lambiase,
\newblock Physics Letters B {\bf 716}, 334 (2012).

\bibitem{Iorio-2013-ID434}
A.~Iorio and G.~Lambiase,
\newblock arxiv:1308.0265 (2013).


\bibitem{dot} G.~Giavaras, P.~A.~Maksym and M.~Roy, J Phys Condens Matter {\bf21} 102201 (2009).
\end{thebibliography}
\end{document}